\documentclass[]{aa}         
\usepackage{natbib}\usepackage{graphicx}      
\bibpunct{(}{)}{;}{a}{}{,}                

\newcommand{\kms}{km\,s$^{-1}$}     
\newcommand{\sqcm}{cm$^{-2}$}  
\newcommand{\lya}{Lyman-$\alpha$}
\newcommand{\lyb}{Lyman-$\beta$}
\newcommand{\lyg}{Lyman-$\gamma$}
\newcommand{\hi}{\ion{H}{i}} 
\newcommand{\hw}{\ion{H}{ii}} 
\newcommand{\os}{\ion{O}{vi}}

\newcommand{\cf}{\ion{C}{iv}}
\newcommand{\siw}{\ion{Si}{ii}}

\newcommand{\sif}{\ion{Si}{iv}}

\newcommand{\nf}{\ion{N}{v}}
\newcommand{\nos}{12}       
\newcommand{\nosass}{three} 
\newcommand{\nosint}{nine}  

\begin{document}

\title{Hot halos around high redshift protogalaxies}
\subtitle{Observations of \os\ and \nf\ absorption in damped \lya\
 systems\thanks{Based on observations taken with the Ultraviolet and
 Visible Echelle Spectrograph (UVES) on the Very Large Telescope (VLT)
 Unit 2 (Kueyen) at Paranal, Chile, operated by ESO.}} 
\author{Andrew J. Fox\inst{1}, Patrick Petitjean\inst{1,2},
 C\'edric Ledoux\inst{3}, \& Raghunathan Srianand\inst{4}}
\institute{Institut d'Astrophysique de Paris, 98bis Boulevard Arago, 75014
  Paris, France; fox@iap.fr \and
  LERMA, Observatoire de Paris, 61 Avenue de l'Observatoire, 75014
  Paris, France \and
  European Southern Observatory, Alonso de C\'ordova 3107, Casilla
  19001, Vitacura, Santiago 19, Chile \and 
  IUCAA, Post Bag 4, Ganesh Khind, Pune 411 007, India}

\date{Received August 1, 2006, Accepted January 12, 2007}

\authorrunning{Fox et al.}
\titlerunning{Hot DLA halos} 

\abstract
{}
{We present a study of the highly ionized gas (plasma)
 associated with damped \lya\ (DLA) systems at $z$=2.1--3.1.}
{We search for \os\ absorption and corresponding \sif, \cf, and \nf\
 in a Very Large Telescope/Ultraviolet-Visible Echelle Spectrograph
 (VLT/UVES) sample of 35 DLA systems with data covering \os\ at
 S/N$>$10. We then use optical depth profile comparisons and ionization
 modelling to investigate the properties, phase structure, and origin
 of the plasma.}
{We report twelve DLAs (\nosint\ intervening and \nosass\ at
 $<$5\,000~\kms\ from the QSO redshift) with detections of \os\ absorption. 
 There are no clear \os\ non-detections, so the incidence of \os\ in
 DLAs is between 34\% (12/35) and 100\%.
 Among these 12 DLAs, \cf\ and \sif\ are seen whenever data
 is available, and \nf\ is detected in 3 cases.
 Analysis of the line widths together with
 photoionization modelling suggests that two phases of DLA plasma exist:
 a hot, collisionally ionized phase (seen in broad \os\ components),
 and a warm, photoionized phase (seen just in narrow \cf\ and \sif\
 components). The presence of inflows and/or outflows is indicated by
 individual \os\ and \cf\ components displaced from the neutral
 gas (either blueshifted or redshifted) by up to 400~\kms.
 We find tentative evidence (98\% confidence) for
 correlations between the DLA metallicity (measured in the neutral
 gas) and high-ion column density, and between the DLA metallicity and
 high-ion line width, as would be expected if supernova-driven
 galactic outflows rather than accretion produced the high ions.
 Using conservative ionization corrections, 
 we find lower limits to
 the total hydrogen column densities in the hot (\os-bearing) and warm
 (\cf-bearing) phases in the range
 log\,$N_{\rm \hw}^{\rm Hot}>19.5$ to $>21.1$, and 
 log\,$N_{\rm \hw}^{\rm Warm}>19.4$ to $>20.9$. 
 On average, the hot and warm phases thus contain $\ga$40\% and $\ga$20\% of
 the baryonic mass of the neutral phase in DLAs, respectively.}
{
 If the temperature in the \os\ phase is $\approx10^6$\,K and so
 $f_{\rm \os}={\rm \os/O}\ll0.2$, the plasma can make a significant
 contribution to the metal budget at high redshift. Additional
 searches for \os\ in Lyman Limit Systems (QSO absorbers with
 $17.0<N_{\rm \hi}<20.3$) will be necessary to determine the total
 quantity of baryons and metals hidden in hot halos at $z\approx2$.} 

\maketitle

\section{Introduction}
Damped Lyman-$\alpha$ (DLA) systems, defined as those QSO absorbers
with \hi\ column densities $N_{\rm \hi}$~$\ge2\times10^{20}$~\sqcm,
represent the largest reservoirs of neutral gas in the redshift range
0--5, and are believed to be the precursors to modern day disk galaxies
\citep[see recent review by][]{Wo05}.
Detailed abundance studies of these systems at various redshifts
\citep{PW02, DZ04, DZ06} have provided a means to trace the process
of cosmic metal enrichment over a large fraction of the age of the Universe.

In addition to the neutral gas phase, 
spectroscopic observations of DLAs have revealed both
cold molecular \citep[$T\la300$\,K;][]{Pe00, Le03, Sr05}
and warm ionized \citep[$T\sim10^4$\,K;][]{Lu96, Le98, WP00a} components,
with a multi-phase structure resembling that seen in the Galactic
interstellar medium (ISM). However, very little is known about the
presence or properties of a hot ($T>10^5$~K)\footnote{There is some
  disagreement in the ISM and IGM literature about 
  the meaning of the terms ``warm'' and ``hot''.
  X-ray astronomers do not consider any gas below $10^6$\,K hot, and
  IGM astronomers now have adopted the term ``warm-hot'' to describe the
  regime with $10^5$ to $10^7$\,K.
  However, throughout this paper we use the ISM traditions of ``warm'' to
  imply $T\sim10^4$\,K and ``hot'' to mean $T>10^5$\,K.}
ionized medium. 
Given the evidence for star formation in DLAs \citep*[see][]{Wo05},
Type II supernovae should create observable regions
of hot, shock-heated interstellar plasma. The separate process of
accretion and shock-heating of intergalactic gas
may also lead to the production of a hot ionized medium. 

The best ultra-violet (UV) lines available for studying hot interstellar plasma
are the \os~$\lambda\lambda1031, 1037$ and \nf~$\lambda\lambda1238, 1242$
doublets. \os\ (\nf) peaks in abundance at $T=3(2)\times10^5$~K under
collisional ionization equilibrium (CIE) conditions \citep{SD93, GS06}.
Surveys of \os\ absorption have been used to establish the presence of a hot
ionized medium in the Milky Way and its surrounding network of
high-velocity clouds \citep{Sa03, Se03, Fo06}.
Additionally, \os\ can be formed by photoionization, as is reported
for narrow \os\ absorbers in the intergalactic medium (IGM) at $z\ga2$
\citep{Ca02, Be02, Be05, Lv03}.
Distinguishing between photoionized and collisionally ionized \os\
absorbers is therefore important in searching for a hot ionized medium in DLAs.

\os\ absorption has been found before in Lyman Limit Systems \citep[QSO
absorbers with $17.0<$~log\,$N_{\rm \hi}<20.3$;][]{KT97, KT99} but no
systematic search for \os\ absorption has yet been conducted in DLAs. In this
paper we present such a search.
In Sect. 2 we discuss the data
acquisition and reduction, \os\ identification, and absorption line
measurements. In Sect. 3 we present the spectra and compare the line
profiles, and in Sect. 4 we
discuss ionization processes. In Sect. 5 and 6 we discuss the total
plasma content in DLAs and its contribution to the cosmic density $\Omega$.
In Sect. 7 we briefly speculate on the origins of the \os\ phase. 
We summarize our study in Sect. 8. 

\section{Data acquisition and handling}

\subsection{Observations}
The VLT/UVES DLA sample
currently consists of 123 DLA (log\,$N_{\rm \hi}$ $\ge$ 20.3) and sub-DLA 
(19.7 $<$ log\,$N_{\rm \hi}$ $<$ 20.3) systems.
All data were acquired with the Ultraviolet-Visible Echelle
Spectrograph (UVES) at the 8.2~m Very Large Telescope Unit 2 (VLT/UT2)
at Paranal, Chile in the years 2000 to 2006. 
The data reduction was performed using 
the interactive pipeline described in \citet{Ba00};
full details of the reduction procedures are given in \citet{Le03}.
The rebinned pixel size is $\approx$2~\kms\ and the data have a
spectral resolution (FWHM) of 6.6~\kms\ ($R$=45\,000).

\subsection{Identifying \os\ absorption in DLAs}
\os\ systems are only accessible from the ground at $z\ga2$,
where the transitions become redshifted enough 
to pass the atmospheric cutoff near 3000~\AA.
They lie in the \lya\ forest, the series of intervening \hi\
absorption lines whose number density 
increases rapidly above $z=2.5$ \citep{Ki97}.
There is therefore a high level of confusion in separating \os\ from \hi\
interlopers, that becomes more difficult with increasing redshift. 
We thus adopted a series of systematic steps to
identify our sample of DLA \os\ absorbers, as follows.

(1) We selected from the UVES DLA sample the 35 DLAs with data
covering the \os\ wavelength range at S/N$>$10 per resolution
element\footnote{Note that 8 sub-DLAs (or ``super-Lyman Limit
  Systems'') were discarded so as to restrict the sample to genuine
  DLAs, even though they have \os\ coverage and adequate S/N.}.

(2) We immediately rejected 19 cases 
where the \os\ lines are clearly blended, e.g. by continuous
absorption troughs extending over several hundred~\kms.

(3) We searched for \os\ components over a range of $\pm$500~\kms\
relative to the redshift of the neutral gas in the DLA.
If absorption lines were found where the doublet ratio
$\tau$(\os\ $\lambda$1031)/$\tau$(\os\ $\lambda$1037)=2 in all 
pixels through the line profile, we made a preliminary identification
of \os. Alternatively, if one \os\ line was blended but the other \os\
line showed a candidate absorption line, we looked at the \cf\ data
(which exists in most cases), and if the candidate \os\ 
absorber showed a similar line profile to the \cf\ absorber,
we treated the \os\ as a preliminary identification.

(4) We investigated whether the preliminary \os\ identifications could be 
caused by intervening \lya, \lyb\ or \lyg\ forest absorbers, by
looking in each case for corresponding absorption in the other Lyman
series lines and in \cf.  If no blending lines were identified, the
preliminary \os\ detections became real and entered our sample.
Four preliminary \os\ identifications were identified as being due to
intervening \hi\ lines. 
We are confident that no \lyb\ or \lyg\ forest
interlopers remain in our sample, since in these cases the corresponding \lya\
would be seen in the spectrum. 
It is more difficult to decontaminate \lya\ absorbers (which are more
numerous) since the corresponding higher order 
Lyman lines often fall outside the observed wavelength range.
Therefore, it is possible that some contamination by \lya\
forest absorbers remains in our data.

The problem of reliably discriminating between true high-redshift \os\
absorbers and intervening Lyman forest systems is discussed in
\citet{BT96} and \citet{Si02}.  
Our situation is favorable in that we look for \os\ at the redshift
$z_{\rm abs}$ where an absorption line system is already known to
  exist, not in a blind search over redshift. 
We can crudely assess the probability that a given \os\ component is
caused by blending using the observed density of \lya\ forest lines.
The data presented in \citet{Ki97} imply a mean separation of
\lya\ forest lines (with log\,$N_{\rm \hi}=13.1-14.3$) of 560~\kms\ at
$z$=2.17--2.45, and of 360~\kms\ at $z$=2.71--2.30. 
We adopt an average separation of 400~\kms\ for the purposes of this
calculation, ignoring the effects of clustering. The probability of a
forest line falling in a $\pm$20~\kms\ interval around 
$z_{\rm abs}$ is $P_{\rm blend}\approx40/400\approx0.1$. 
The probability of two blending lines falling at the same velocity,
one in \os\ $\lambda$1031 and the other in $\lambda$1037, or
equivalently two blends falling in one \os\ line at the velocity of 
two known components, is $P_{\rm blend}^2\approx0.01$, 
and so on for multiple lines. 
We find several DLAs with multiple \os\ components identified; there
is a negligible probability that they are all blends. In other
cases there are only one or two components identified; these
detections are less secure.
To further justify our identification of \os\ in each
DLA, we present detailed notes on each system in the appendix, which
should be consulted together with Figure 1, showing the spectra.

\subsection{The resulting sample}
After applying our search criteria we are left with \nos\ \os\
detections, of which three are at 
d$v$=$c|z_{\rm qso}-z_{\rm abs}|/(1+z_{\rm abs})<5\,000$~\kms\ from the
QSO redshift, and so may be ``associated'' by the traditional definition
(though see Sect. 5.1). 
\cf\ and \sif\ are seen in every case where data is
available, and \nf\ is detected in 3/9 systems with data. 
The \os\ detections cover the redshift range 2.07--3.08, median
redshift 2.62, 
and are observed at S/N levels (per resolution element) in the range 
23--41 at the \os\ wavelength. 
The S/N is typically over twice as high near \cf\ and \sif.  
The DLAs span values of log\,$N_{\rm \hi}$ from 20.35 to 21.75, with
metallicities [Z/H]$_{\rm Neut}$
(measured from the singly-ionized species in the neutral phase)
also spanning two orders of magnitude between $-$2.59 and $-$0.49
\citep{Le06}\footnote{Throughout this paper we use the convention 
  [Z/H]=log($N_{\rm Z}/N_{\rm H}$)--log($N_{\rm Z}/N_{\rm H}$)$_{\odot}$.}.
The three systems with $z_{\rm abs}\approx z_{\rm qso}$ are toward
Q0528--250, Q0841+129, and Q2059--360. 

\subsection{Incidence of \os\ absorption in DLAs}
We detect \os\ in 12 of 35 DLAs (34\%) with \os\ coverage. 
There are no clear \os\ non-detections, so the percentage of DLAs with
\os\ could be 100\%. However, we caution that \os\ non-detections are
difficult to make, due to the high density of forest lines; 
a conservative estimate of the fraction of DLAs with \os\ is thus $>$34\%. 
Analyzing only the intervening DLAs at $>$5\,000~\kms\ from
the quasar redshift, we find that 9 of 31 show \os, so the incidence
of \os\ in intervening DLAs is $>$29\%, whereas 3/4
DLAs at $<$5\,000~\kms\ from the quasar redshift show \os. 

\subsection{Measurement of absorption}
For each DLA we determined 
by eye the velocity range of absorption (labelled $v_-$ to $v_+$) in 
\os~$\lambda\lambda1031, 1037$, \sif~$\lambda\lambda1393, 1402$,
\cf~$\lambda\lambda1548, 1550$, \nf~$\lambda\lambda1238, 1242$, and
the neutral gas\footnote{
Note that the ``neutral'' phase is traced here
through absorption lines from singly-ionized metals; it is so called
because hydrogen is predominantly neutral in this phase.}
The exact range of ions available changes
from system to system, due to blending and data availability. 
Our neutral-phase line for each DLA is adopted from \citet{Le06}, who chose
a line with $0.1<F(v)/F_c(v)<0.6$ in the strongest component to
minimize saturation, where $F_c(v)$ and $F(v)$ 
are the continuum and observed fluxes as a function of velocity.

We fit continua locally to all
the lines of interest using low-order (often linear) Legendre 
polynomials to regions of continuum free from absorption.
We then integrated the absorption in each line using the apparent
optical depth (AOD) method of \citet{SS91}, which returns measurements
of the column density
$N_{\mathrm{a}}=[3.768\times10^{14}/(\lambda_0
  f)]\int_{v_-}^{v_+}\tau_a(v){\mathrm 
  dv}$~\kms, where $\tau_a(v)=\mathrm{ln}[F_c(v)/F(v)]$, $\lambda_0$
is the rest wavelength in Angstroms, 
and $f$ is the oscillator strength of the line \citep[taken from][]{Mo03}.
The AOD method also returns measurements of the central velocity $\bar{v}$
of the absorption using the
first moment of the optical depth profile,
$\bar{v}=\int^{v_+}_{v_-}v\tau_a(v)\mathrm{d}v/\int^{v_+}_{v_-}
\tau_a(v)\mathrm{d}v$.
The treatment of errors arising from statistical (photon noise)
and systematic (continuum placement) effects is handled as described
in \citet{SS92}. 
The UVES instrumental resolution of 6.6~\kms\ (FWHM)
ensures that lines with $b>4$~\kms\ will be resolved,
which is required for the AOD method to give accurate results.

For each line we calculate $\Delta v$, the velocity range containing
the central 90\% of the total integrated optical depth in the line
\citep{PW97b}. For cases where no absorption is seen at the DLA
redshift in \nf, we present a 3$\sigma$ upper limit to the column
density, using the 3$\sigma$ upper limit to the equivalent width, and
assuming any absorption to be optically thin. For cases
where the line profile is saturated (which we define as where
$F(v)-\sigma_{F(v)}<0$ in any pixel within the line profile), we
present a lower limit to the column density,
and an upper limit to the line width.

We also used the VPFIT software
package\footnote{Available at http://www.ast.cam.ac.uk/$\sim$rfc/vpfit.html}  
to derive the absorption line parameters for each high-ion absorber
with Voigt profile fits.
This provides an independent check of our absorption line measurements.
The number of components fit is somewhat subjective;
we use the minimum number necessary.
We present the measurements using both the AOD and VPFIT techniques in
Table 1. Note that the AOD method, which can be fully automated for high
efficiency, and has the advantage of not requiring any assumptions
about the component structure, generally reproduces the same column
density results (within the errors) as the VPFIT method. 
This indicates that the choice of $v_-$ and $v_+$ by eye is not a
significant source of error.

All velocities (in Table 1 and throughout this paper) are quoted
relative to the redshift of the neutral gas in the DLA, which is
measured using the central velocity of the strongest component of the
low-ionization metal absorption in \citet{Le06}. 
This provides a convenient rest-frame
in which to discuss the kinematics of the absorption line profiles.

\section{Results}
In Fig. 1 we show the absorption line profiles of all the DLA
systems with \os\ detections. Fig 2. shows the $N_{\rm \cf}$/$N_{\rm
  \os}$ ratios versus velocity for the ten DLAs in our sample with
both \os\ and \cf\ data.

\begin{figure*}
\centering
\includegraphics[width=17cm]{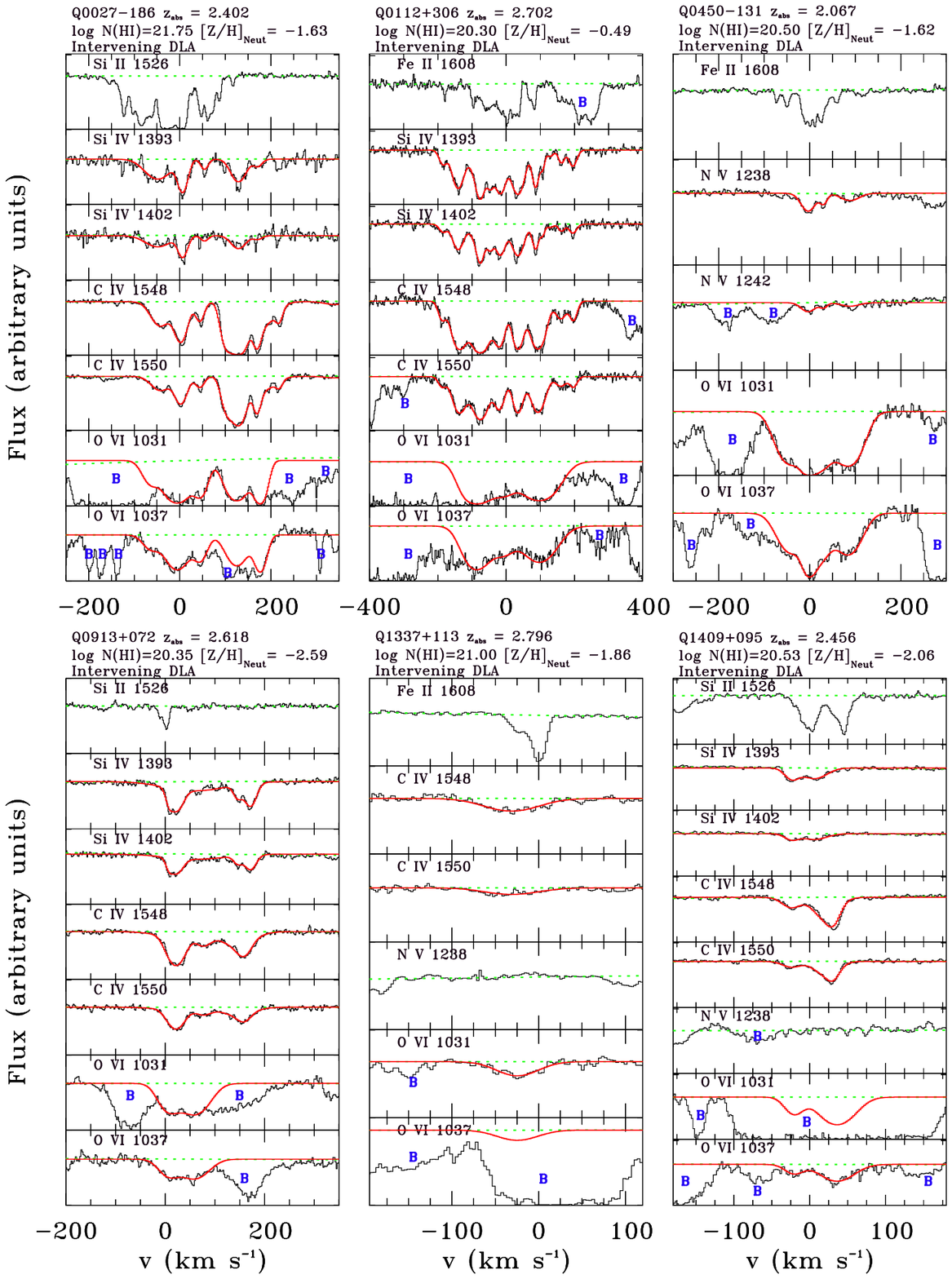}
\caption{VLT/UVES absorption line spectra of all DLA systems with
  detections of \os\ absorption.
  The tracer of the neutral gas is shown in the top panel, with the
  other panels showing all available high-ion data.
  In each DLA $v=0$~\kms\
  corresponds to the strongest component of absorption in the neutral gas.
  The red line shows our VPFIT model of the absorption, and
  fitted continua are shown as light dashed lines. 
  Blends are identified with the letter 'B'. 
}
\end{figure*}
\addtocounter{figure}{-1}
\begin{figure*}
\centering
\includegraphics[width=17cm]{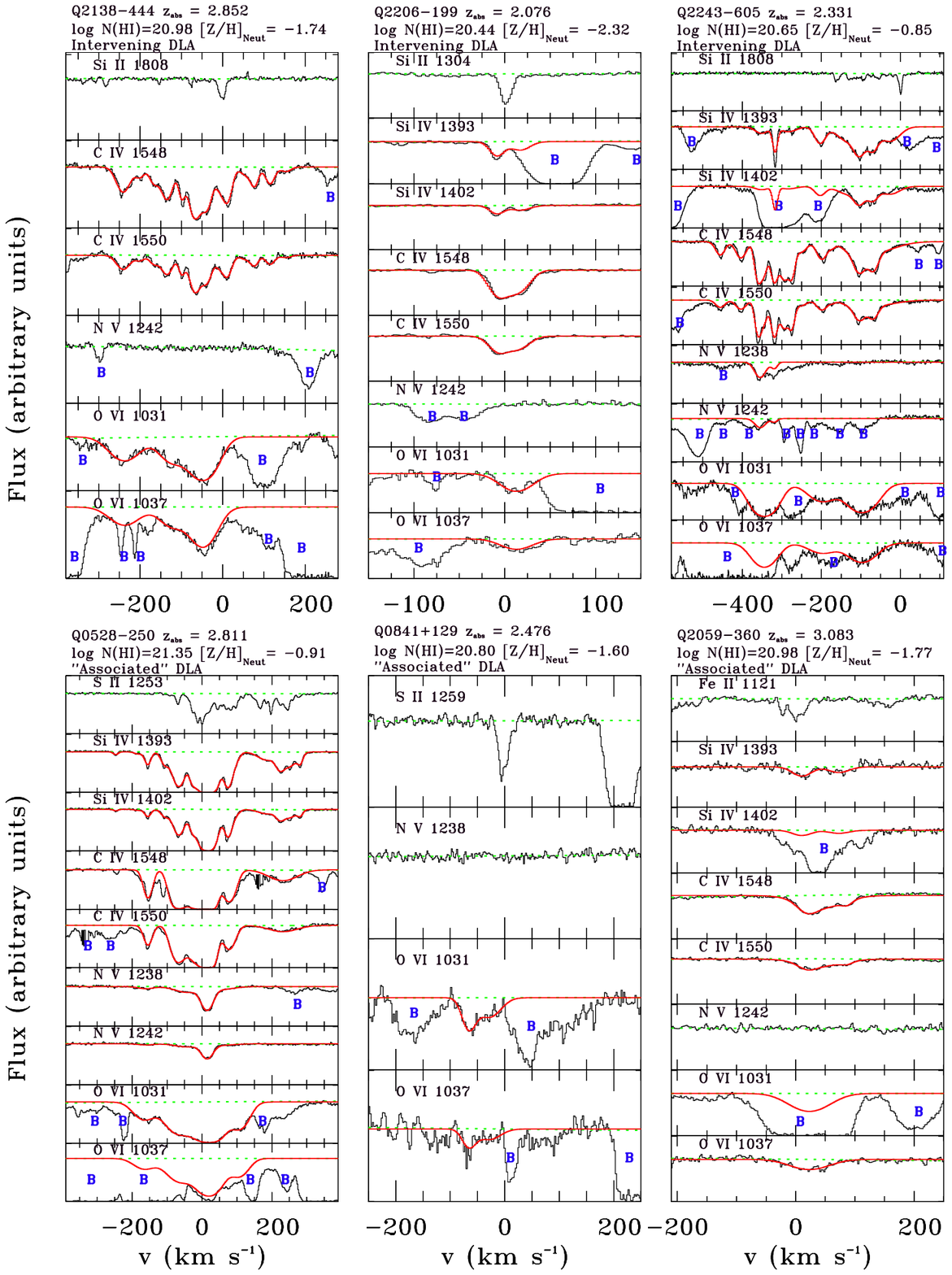}
\caption{(-continued). All DLA systems with \os\
  absorption. The lower three systems are at $<$5\,000~\kms\ from the QSO.}
\end{figure*}

\begin{figure}
\resizebox{\hsize}{!} 
{\includegraphics[width=15cm]{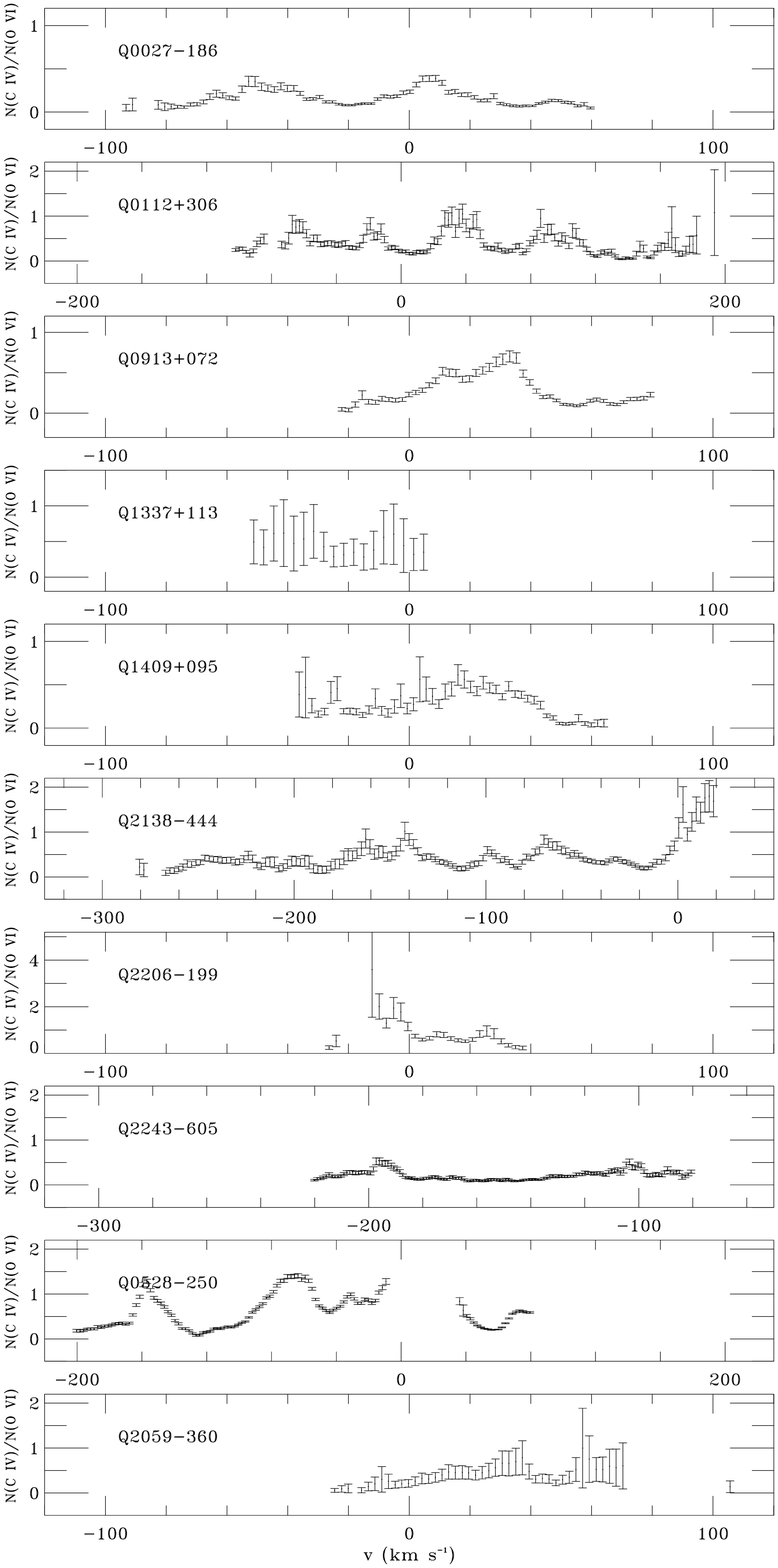}}
\caption{
  $N_{\rm \cf}$/$N_{\rm \os}$ ratio versus velocity. 
  Each data point on this plot is a velocity pixel where both \cf\ and
  \os\ are detected and unsaturated.
  $N_{\rm \cf}$ is derived from the optical depth profile of
  $\lambda$1548 unless this line is saturated, when we use $\lambda$1550. 
  $N_{\rm \os}$ is derived from $\lambda$1031 unless 
  this line is blended, when we use $\lambda$1037.
  In nine out of ten cases (the exception is the DLA toward Q1337+113), this
  ratio is non-linear with velocity, showing that the \os\ and \cf\
  ion are generally not co-spatial.
}
\end{figure}

There is considerable variation in the appearance of the highly ionized 
absorption lines in the DLAs. The \os\ absorbers range from 
cases with a single, optically thin component
(e.g. toward Q2059-360 or Q2206-199) to cases with a series of saturated
components (e.g. toward Q0450-131 or Q2243-605).
The \cf\ profiles range from cases with one or two components spanning
$<$100~\kms\ to cases with $\ga$15 components spanning several
hundred~\kms. By no means is it clear that we are looking at a
homogeneous sample of objects.

\subsection{\os\ column density} 
The logarithmic integrated \os\ column density in our \nos\ detections
ranges from 13.66 to $>$15.15, with a median value of 14.77, a mean of
14.54, and a standard deviation of 0.59. 
The mean column density is similar 
to that seen in the ``thick disk'' of \os\ in the
Milky Way \citep{Sa03}, where log\,$N_{\rm \os}$ ranges from 13.85 to
14.78, with a median value of 14.38.
A typical complete sight line through the Milky Way halo also shows \os\
high-velocity clouds \citep*[HVCs;][]{Se03, Fo06}, so a better
analogy for DLAs is the total \os\ column density 
through the Milky Way halo over all velocities, which is typically
14.80. 
The average value of $N_{\rm \os}$ in DLAs is slightly larger 
than both the Large Magellanic Cloud average of
$\langle$log\,$N_{\rm \os}\rangle$=14.37 \citep{Ho02}, 
and the amount log\,$N_{\rm \os}$=14.3 seen in starburst galaxy
NGC~1705 \citep{He01}.
The values of $N_{\rm \hi}$ in the neutral gas in the DLAs, between
20.35 to 21.75, 
compare to values in the Galactic ISM of between 19.6 and $>$22.4
\citep{DL90}, with over two orders of magnitude of dispersion in each
case. The similarity in the amount of \os\ seen is intriguing
considering that the metallicities in these DLA systems 
are typically forty times lower 
than the ($\sim$solar) value in the Galactic ISM. 
This implies that the total column density of ionized gas in DLAs is
very large (see Sect. 5).

\subsection{Comparison of high ion line profiles}
Throughout this Section we refer to 
Table 2, comparing the integrated kinematic measurements of the
neutral-phase, \cf, and \os\ absorption line profiles,
Fig. 3, showing the column density, line width, and central velocity
distributions for the components of \sif, \cf, and \os\ in the DLA
sample, and Fig. 4, comparing many measured absorber properties using
scatter plots. 

\begin{figure}
\resizebox{\hsize}{!} 
{\includegraphics[width=15cm]{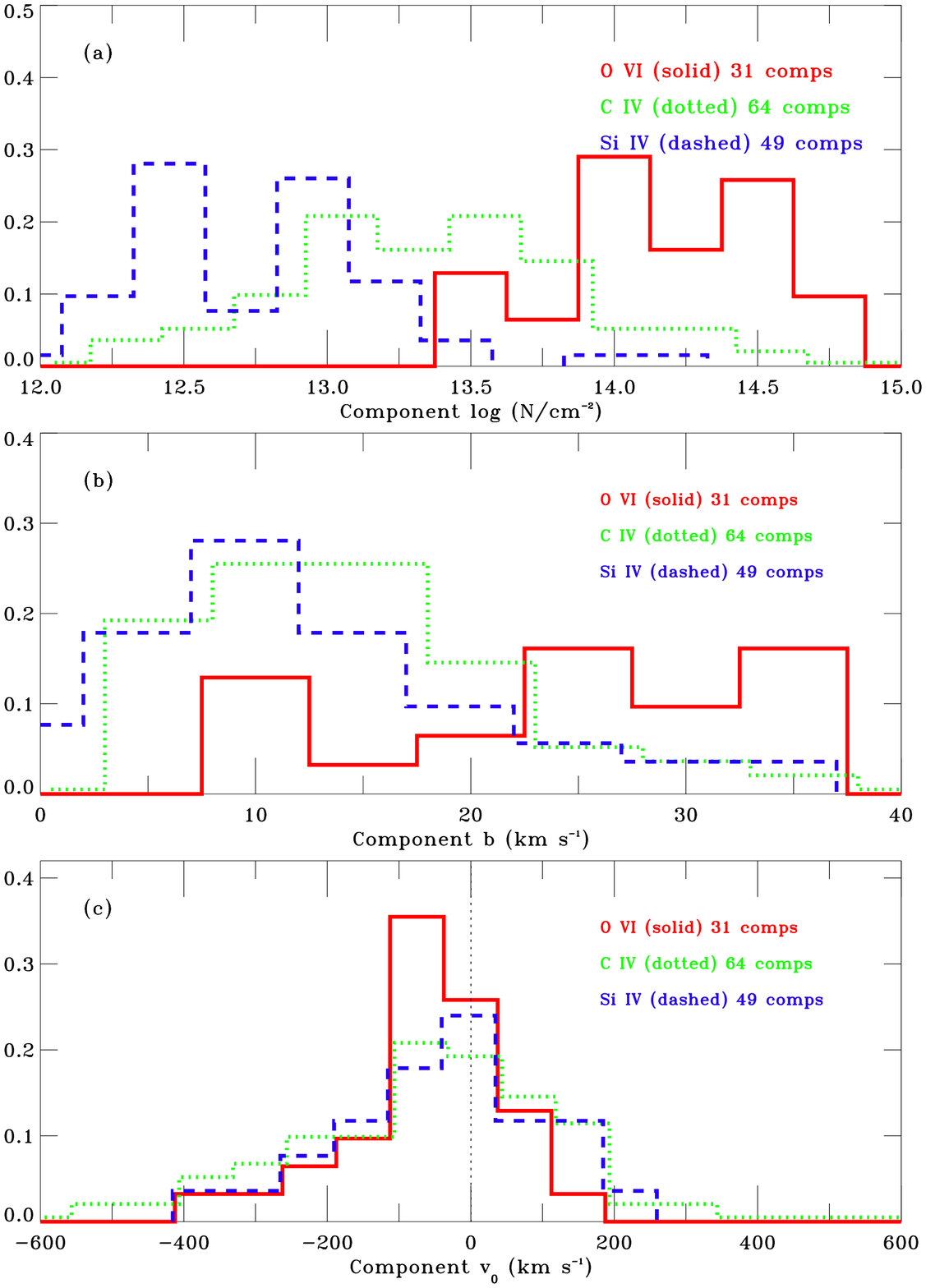}}
\caption{
  Normalized histograms of the properties of high-ion components that
  comprise the DLA absorbers, as measured using VPFIT.
  The number of components in each sample is indicated on the plot.
  Panel (a): Column density distributions.
  Panel (b): Line width distributions.
  Note that
  $\bar{b}_{\rm \os}>\bar{b}_{\rm \cf}>\bar{b}_{\rm \sif} $, i.e. the
  average component width rises with ionization potential.
  Panel (c): Central velocity distributions. Note the large dispersion in
  individual component velocities, at up to 400~\kms.
}
\end{figure}

\begin{figure*}  
\centering       
\includegraphics[width=19cm]{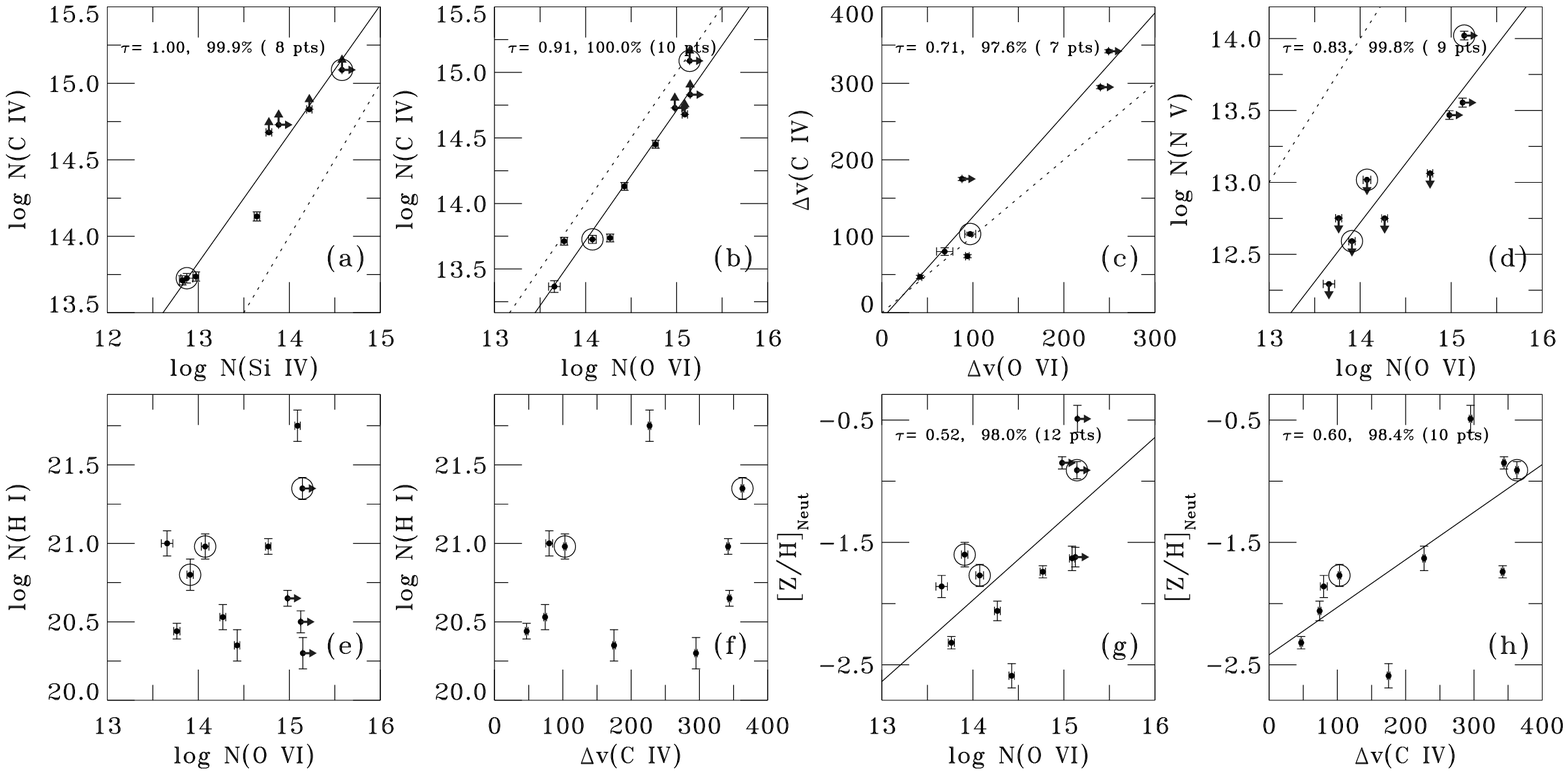} 
\caption{
  Comparisons between measured properties of \os\ DLAs. 
  $N$ is the total column density over all components, 
  in units of \sqcm; 
  $\Delta v$ is in \kms.
  When a positive correlation between the two variables plotted is found 
  (treating the upper/lower limits as data points),
  we display the Kendall rank correlation coefficient $\tau$ and 
  its significance  
  on the panel, together with a least-square linear fit to the data
  (solid line). Dotted lines show one-to-one relationships.
  Saturated absorbers are shown with lower limits to log\,$N$. 
  The three DLAs at $<$5\,000~\kms\ from the quasar redshift are
  highlighted in circles.
  }
\end{figure*}

The properties of the \sif\ and \cf\ absorbers are strongly correlated. 
This is revealed by visual examination of the
data, by the close similarity in their distributions of individual
component line width and central 
velocities in Fig. 3, and by the strong correlation in
their total column densities (Fig. 4, panel a). 
In general the \sif\ and \cf\ thus arise in the same gaseous phase.

There is a $>$3$\sigma$ correlation between $N_{\rm \os}$ and $N_{\rm
\cf}$ (Fig. 4, panel b). The total line widths of the two ions are
also correlated (Fig. 4, panel c). Despite these correlations in the
integrated properties, we do observe significant differences in the \cf\
and \os\ absorbers at the component level. 
The \cf/\os\ column density ratios are non-linear with velocity in
nine out of ten cases where data on both ions exist, 
often showing multiple peaks and troughs within the
velocity width of each system (Fig. 2). This can be attributed to 
the \os\ profiles being smoother than those of \cf, with fewer
\os\ components overall and no narrow \os\ components observed. 
These non-linear ratios imply the \cf\ and \os\ arise in separate phases.
Further evidence for \os\ and \cf\ tracing
different gas phases in DLAs is given by the histograms of the line
widths of the individual components (Fig. 3, panel b), 
where strong differences are seen between \cf\ and \os.
Many narrow ($b\la10$~\kms) \cf\ components are observed, but most \os\
components have $b$-values in the range 25--40~\kms. 

We report three DLAs with detections of both \os\ and \nf. 
One of these DLAs is close to the QSO redshift (toward Q0528-250)
leaving two intervening cases 
where \nf\ is seen (toward Q0450-131 and Q2243-605). 
The \os\ column densities are highest (in fact, saturated) in the
cases where \nf\ is present (Fig. 4, panel d). 
Profile comparisons are hence
not useful in revealing the \os-to-\nf\ relationship. 

\subsection{High ions versus neutral species}
In Fig. 5 we compare the velocity range $v_-$ to $v_+$ between \os, \cf,
and the neutral lines in each DLA, to display the larger range of
velocity space that the ionized lines invariably occupy. 
The mean total width (and its standard deviation) of all the \cf\
absorbers (including the limits from the saturated cases),
$\langle \Delta v_{\rm \cf}\rangle$=205$\pm$125~\kms, 
is twice as broad as the average width of the neutral lines, 
$\langle \Delta v_{\rm neut}\rangle$=102$\pm$88~\kms. 
These results agree with the corresponding numbers reported in
\citet*{Wo05} over a sample of over 70 DLAs: 
$\langle \Delta v_{\rm \cf}\rangle$=209$\pm$114~\kms\
and $\langle \Delta v_{\rm neut}\rangle$=114$\pm$84~\kms. 
 
\begin{figure}
\resizebox{\hsize}{!} 
{\includegraphics{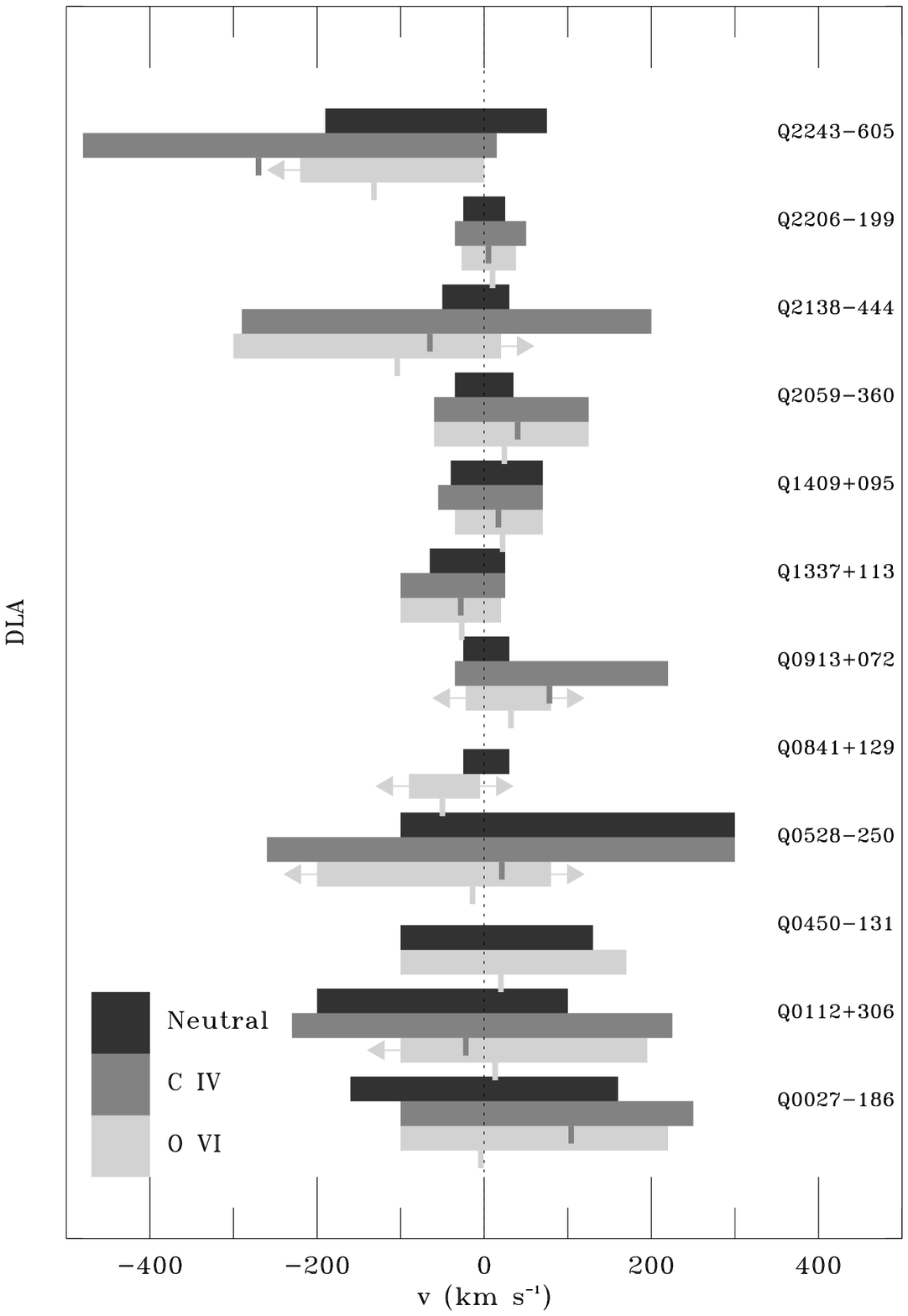}}
\caption{
  Velocity range of absorption $v_-$ to $v_+$ for \os, \cf, and
  the neutral gas in each DLA, showing the larger velocity dispersion
  in the ionized phases.
  Tick marks indicate the mean velocity of the
  \cf\ and \os\ absorption. 
  Partial blending limits the velocity range where we can
  measure the \os\ absorption in several cases, marked with arrows.
  The range $v_+-v_-$ is always larger than
  $\Delta v$, our primary measure of line width; it is used here to
  show the distribution of ionized gas around the DLA redshift.}
\end{figure}

Individual \cf\ components are found at velocities of up to 400~\kms\ 
from the neutral gas (see the distribution in Fig. 3, panel c). 
We also observe offsets in the overall mean velocity $\bar{v}$ of high-ion
absorption in each DLA.
The mean and standard deviation of $|\bar{v}_{\rm \cf}|$ is 65$\pm$78~\kms. 
These offsets can be interpreted as the signature of galactic inflows
and/or outflows. 

In Fig. 4 (panels e to h) 
we directly compare our new measured high-ion DLA properties with the
\hi\ column densities and neutral-phase metallicities.
We find that the column densities and line widths of the high ions are each
\emph{uncorrelated} with the column density of \hi\ in the neutral
gas. However, the plasma properties do
depend on the metallicity of the neutral gas.
This is revealed by the tentative detection (98\% confidence) of 
a [Z/H]$_{\rm Neut}$ vs $N_{\rm \os}$ correlation (panel g),
and the tentative detection (also at 98\% confidence)
of a [Z/H]$_{\rm Neut}$ vs $\Delta v_{\rm \cf}$ correlation (panel h). 
Note we use $\Delta v_{\rm \cf}$ and not
$\Delta v_{\rm \os}$ to illustrate the high-ion kinematics since
\cf\ line widths are far better measured. 
All DLAs in the sample with [Z/H]$_{\rm Neut}$ $>-$1.5 have both
log\,$N_{\rm \os}>$14.77 (the sample median value) and 
$\Delta v_{\rm \cf}>$250\kms. 
A similar correlation is found between [Z/H]$_{\rm Neut}$ and 
$N_{\rm \cf}$ (not shown).

The correlation between the high-ion line width and metallicity in DLAs
follows the observed correlation between the {\it low-ion}
line width and metallicity \citep{Le06, WP98}.
However, these correlations likely have very different
explanations. 
The low-ion velocity dispersion is thought to be
dominated by gravity, and so its correlation with metallicity may
imply an underlying mass-metallicity relation \citep{Le06}.
The high-ion line width may be correlated with metallicity since
both high ions and metals can be produced as a result of supernovae.
Further observations of DLA high ions are needed to
investigate the existence of this correlation, which in principle can
be used to probe the origin of plasma in DLAs.

\section{Ionization}   
To create the ions \os, \nf, \cf, and \sif\ from the preceding
ionization stages requires 114, 78, 48, and 34~eV, respectively.
In principle, these ionization energies can be supplied either by
electron collisions in hot plasma or by extreme-UV photons. 
Either ionization process is plausible in the vicinity of DLAs, 
due to the local presence of type II supernovae, accreting
intergalactic gas, and the ionizing extragalactic background.

\subsection{Model-independent insights}
Many of the \cf\ and \sif\ components have narrow line widths, $b\la10$~\kms\
implying photoionization is the origin for these components. 
Collisional ionization is ruled out since $b_{\rm \cf}<10$~\kms\ implies
log\,$T<4.86$, 
at which temperature little \cf\ would be produced
by collisions with electrons \citep{SD93}; a similar argument holds for \sif.
These narrow components are particularly noticeable in
the DLAs toward Q2243-603, Q0112-306, and Q2138-444.
There are also broader \cf\ components present in the data with
$b\approx30-40$~\kms, e.g. toward Q1337+113 and Q2059-360.
However, no individual narrow photoionized \os\ components are found;
the narrowest \os\ $b$-value is 14~\kms, with the majority of
cases over 20~\kms.
For comparison, a thermally broadened \os\ component at a temperature of 
$3\times10^5$~K, where \os\ peaks in abundance in collisionally
ionized plasma, has a line width $b_{\rm \os}=18$~\kms.
The line widths of the \nf\ components are generally over 20~\kms,
though the sample is small.
We cannot infer conclusively from the observed line widths whether
\os\ and \nf\ are photoionized or collisionally ionized, because of
the possibility of non-thermal broadening or unresolved components.
{\it However, the fact that we clearly see photoionized \cf\ and \sif\
components but do not see these components in \os\ suggests that the
\os\ arises in a separate, collisionally ionized phase.}
Analogous situations (with narrow components seen in \cf\ but not
in \os) are observed in the low halo of the Milky Way
\citep{Fo03} and in Lyman Limit Systems \citep{KT99}.

\subsection{Photoionization models}
We ran photoionization models using the code CLOUDY 
\citep[version C06.02, described in][]{Fe98}, 
to investigate the viability of photoionization as an origin mechanism
for the observed high ions in the DLA at $z_{\rm abs}=2.076$ toward Q2206-199.
This absorber was chosen as an illustrative case.
Our methodology involved running a grid of models at different 
gas density and total hydrogen column density, until we found a
solution that reproduced the high-ion column densities (see Fig. 6).

The following assumptions were made:
(1) the plasma in DLAs exists in a uniform, plane-parallel slab; 
(2) the plasma has the same metallicity as the neutral gas; 
(3) the plasma has solar relative abundance ratios; 
(4) the plasma is exposed to the extragalactic background taken
from \citet{HM96}, with a mean intensity at 912~\AA\ of
$J_{912}=10^{-21.5}$~ergs~\sqcm~s$^{-1}$~Hz$^{-1}$~sr$^{-1}$
at $z=2$.

\begin{figure*}
\centering   
{\includegraphics[width=15cm]{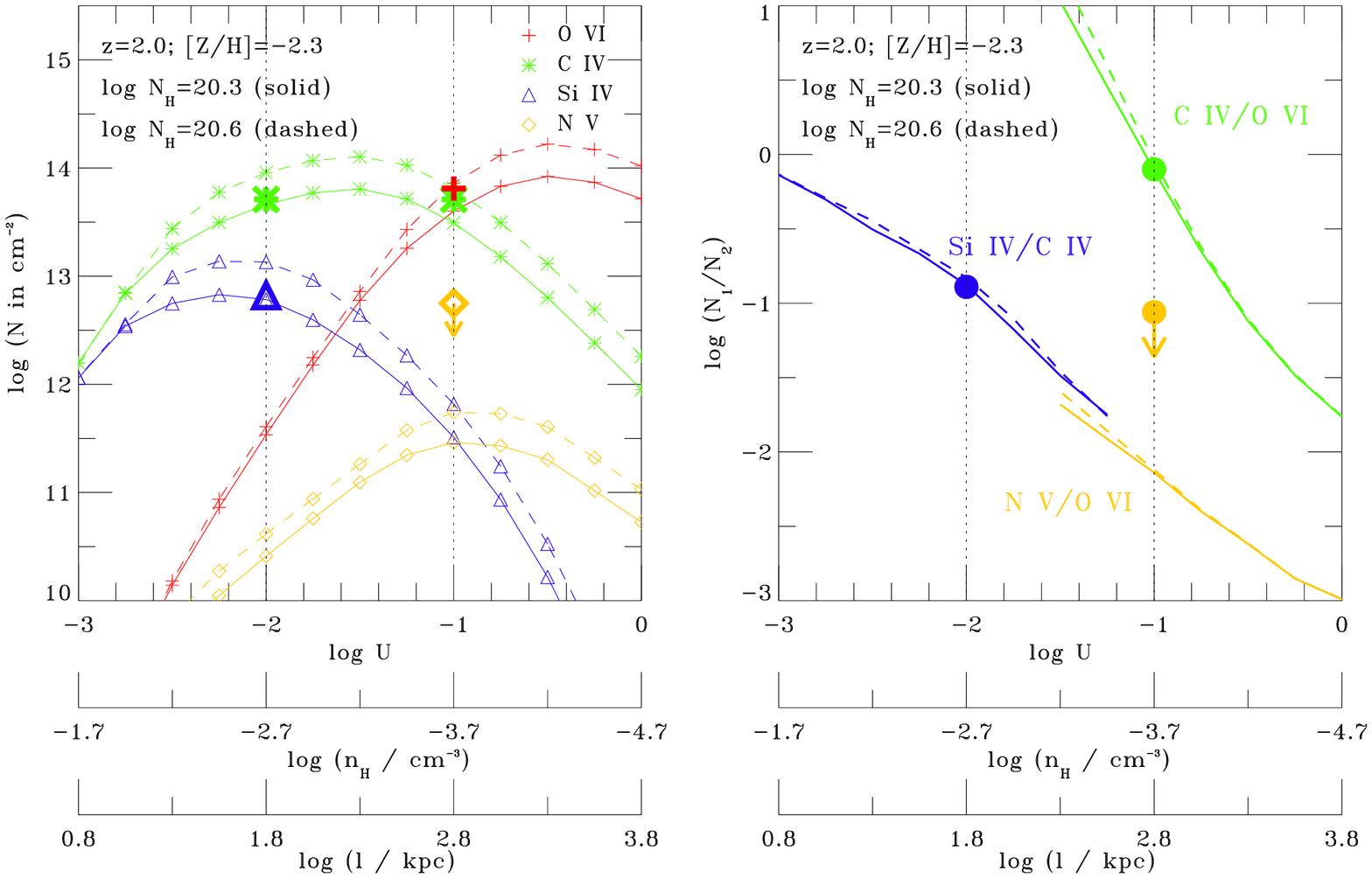}}
\caption{Investigation of whether photoionization by the extragalactic
  background can explain the high ions in the DLA at
  $z_{\rm abs}=2.076$ toward Q2206-199, 
  assuming solar relative abundances. 
  The high-ion column densities ($N$, left panel) and their
  ratios ($N_{1}$/$N_{2}$, right panel) are plotted as
  a function of ionization parameter $U$, the ratio of ionizing photon
  density to gas density, as determined using a series of CLOUDY runs.
  The observations are shown as large symbols. 
  We find the \cf\ and \sif\ observations can
  be reproduced by a model with total hydrogen column density 
  log\,$N_{\rm H}$=20.3 and 
  gas density log~$n_{\rm H}=-2.7$. This solution is physically reasonable.
  The \cf\ and \os\ observations can be reproduced by a model with
  log\,$N_{\rm H}$=20.6 and log~$n_{\rm H}=-3.7$. 
  However, this solution implies a cloud size of $\approx$630~kpc; 
  we rule this out since the Hubble broadening of such an absorber
  would be broader than the observed \cf\ and \os\ component widths. 
  [O/C]$>$0 is required to produce a
  solution for \cf\ and \os\ that is not unreasonably large;
  collisional ionization models are also possible for \os. 
  The log\,$l$ axis only applies to the case log\,$N_{\rm H}$=20.6. 
}
\end{figure*}

The first result is that in this test case (the DLA toward Q2206-199),
there is no single-phase solution to the
\sif, \cf, and \os\ observations.
There are solutions just to the \cf\ and \sif, or to the \cf\
and \os\ data, which we discuss here. 
A model to the observed $N_{\rm \sif}$ and $N_{\rm \cf}$,
assuming [Si/H]=[C/H],
requires log\,$N_{\rm H}=20.3$ and log\,$U\approx-2.0$, 
where $U$ is the ratio of ionizing photon density to gas density,  
corresponding to log\,$n_{\rm H}=-2.7$, 
or an over-density $\delta=\rho/\bar{\rho}\approx200$.
The \cf\ and \os\ data are reproduced, if we assume [C/H]=[O/H],
by a model with log\,$N_{\rm H}=20.6$, 
ionization parameter log\,$U=-1.0$, 
gas density log\,($n_{\rm H}$/cm$^{-3}$)=$-$3.7, 
and linear size $l=630$~kpc. 
However, using a value for the Hubble constant at $z=2.5$ of
$H_{2.5}$=220~$h_{70}$~\kms~Mpc$^{-1}$ that follows from a 
standard ($\Omega_\Lambda$=0.7, $\Omega_M$=0.3) cosmology, 
the Hubble flow broadening $b$(\kms$)=H_{2.5}L$(Mpc) over a path 
of 630~kpc would create lines with widths of $\approx$140~\kms, 
several times larger than the observed \cf\ line width (see
spectra in Fig. 1).
In order to arrive at a physically reasonable size, one requires
[O/C]$>$0, as is the case for metal-poor halo stars \citep{Ak04}.
In this case a given $N_{\rm \cf}$/$N_{\rm \os}$ ratio can be 
reached at a lower ionization parameter, corresponding to higher
density and smaller size.

If local galactic sources of radiation were present in addition to the
extragalactic background \citep[e.g.][]{HS99, Vl01}, then 
(assuming the shape of the radiation field does not change) 
a given ionization parameter would be
reproduced at higher density and smaller cloud size, and
consequently the Hubble broadening would be smaller.
However, one would not expect a strong flux of \os-ionizing photons
from galactic sources of radiation \citep[e.g.][]{Fo05}, so 
though local sources of radiation could produce the \cf\ they could
not easily explain the \os.

\nf\ is not detected in the Q2206-199 DLA, in agreement with the model
prediction. 
For those DLAs where \nf\ is present, one cannot use the \nf/\cf\
ratios to derive the gas density since the relative nitrogen abundance
in DLAs can be highly non-solar \citep{Pe02}, and also because the
\nf\ may be collisionally ionized. 
We also note that two of the three DLAs at $<$5\,000~\kms\ from the
QSO (toward Q0841+129 and Q2059-360) {\it do not} show
\nf\ absorption, whereas \nf\ is expected in intrinsic systems
photoionized by radiation from the QSO \citep{Ha97}.
This suggests that the influence of photoionizing QSO radiation
is not that strong in these two cases. Indeed, the ionization properties
of the three $z_{\rm abs}\approx z_{\rm qso}$ DLAs are no different than
those of the intervening DLAs, so they may be representative of the
whole population.

\section{Total ionized column density}
We can calculate the total hydrogen column density in the hot
(\os-bearing) and warm (\cf-bearing) phases using 

\begin{equation} 
N_{\rm \hw}^{\rm Hot}
=\frac{N_{\rm \os}}{f_{\rm \os}{\rm (O/H)_{Hot}}}
=\frac{N_{\rm \os}}{f_{\rm \os}{\rm (O/H)_{Neut}}}
 \frac{\rm (O/H)_{Neut}}{\rm (O/H)_{Hot}}
\end{equation}

\begin{equation} 
N_{\rm \hw}^{\rm Warm}
=\frac{N_{\rm \cf}}{f_{\rm \cf}{\rm (C/H)_{Warm}}}
=\frac{N_{\rm \cf}}{f_{\rm \cf}{\rm (C/H)_{Neut}}}
 \frac{\rm (C/H)_{Neut}}{\rm (C/H)_{Warm}}
\end{equation}

where $f_{\rm \os}=N_{\rm \os}$/$N_{\rm O}$ is the fraction of
oxygen atoms in the five-times-ionized state, 
and $f_{\rm \cf}=N_{\rm \cf}$/$N_{\rm C}$.
Assuming solar relative abundance ratios in both the neutral and plasma
phases, we have [O/H]$_{\rm Hot}$=[Z/H]$_{\rm Neut}$ and 
[C/H]$_{\rm Warm}$=[Z/H]$_{\rm Neut}$. By using 
$\rm Z_{Hot}$, $\rm Z_{Warm}$, and 
$\rm Z_{Neut}$ as shorthands for the absolute metallicities in the
hot, warm, and neutral phases, we can write

\begin{equation} 
N_{\rm \hw}^{\rm Hot}
=\frac{N_{\rm \os}}{f_{\rm \os}10^{\rm [Z/H]_{Neut}}
 ({\rm O/H})_\odot}\frac{\rm Z_{Neut}}{\rm Z_{Hot}},
\end{equation}
\begin{equation} 
N_{\rm \hw}^{\rm Warm}
=\frac{N_{\rm \cf}}{f_{\rm \cf}10^{\rm [Z/H]_{Neut}}({\rm
    C/H})_\odot}\frac{\rm Z_{Neut}}{\rm Z_{Warm}},
\end{equation}

To evaluate Eqns. (3) and (4) for each DLA we take 
(O/H)$_\odot$=10$^{-3.34}$ from \citet{As04} and 
(C/H)$_\odot$=10$^{-3.61}$ from \citet*{AP02}, and [Z/H]$_{\rm Neut}$
from \citet{Le06}.
We cannot directly measure the metallicity of the plasma phases 
so we assume 
${\rm Z_{Neut}}/{\rm Z_{Hot}}$ and
${\rm Z_{Neut}}/{\rm Z_{Warm}}$ to be
equal to one in our calculations.
The remaining unknowns are $f_{\rm \os}$ and $f_{\rm \cf}$, which will
depend on the ionization mechanism.
The \emph{maximum} value $f_{\rm \os}$ reaches in collisionally
ionized plasma is 0.2 at $10^{5.45}$~K \citep{SD93}. 
The maximum for $f_{\rm \cf}$ in collisionally ionized plasma is 0.29 at
$10^{5.00}$~K, and our photoionization models also show a maximum
$f_{\rm \cf}$=0.3 in photoionized plasma. 
We can therefore estimate \emph{lower} limits to the hot and warm hydrogen
column densities using the maximum values for $f_{\rm \os}$ and $f_{\rm \cf}$.

The implied hot hydrogen column densities are compared with 
the \hi\ column density in the neutral phase in Table 3 and Figure 7,
where we find 
log\,$N_{\rm \hw}^{\rm Hot}$ ranges from $>$19.5 to $>$21.1, and 
log\,$N_{\rm \hw}^{\rm Warm}$ ranges from $>$19.4 to $>$20.9. 
These lower limits are typically on the same order as the \hi\ column
in the neutral gas, though we observe a considerable dispersion (over
two orders of magnitude) in the value of $N_{\rm \hw}/N_{\rm \hi}$.
Note that if some \cf\ co-existed with the
\os\ in the hot phase, as is possible in collisional ionization models,
then the true hydrogen column in the {\it warm} phase would be lower than that
calculated here. Therefore we do not sum the \hw\ contents of the \cf\
and \os\ phases, since this could constitute a double-counting problem.

\begin{figure}
\resizebox{\hsize}{!} 
{\includegraphics{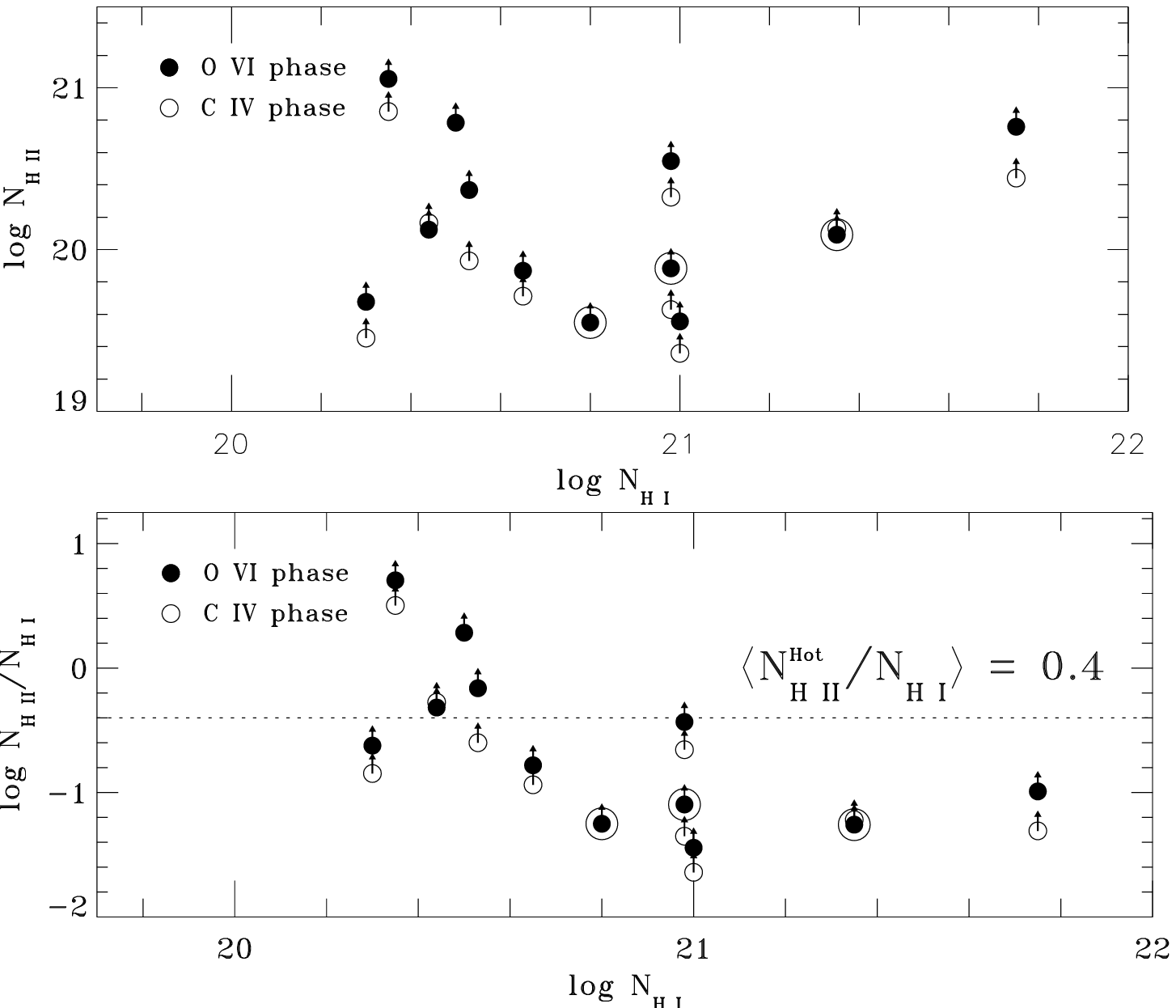}}
\caption{
  Comparison of the total ionized and neutral column densities in DLAs.
  The top panel directly compares the total ionized column
  density $N_{\rm \hw}$ in both the hot and warm phases to $N_{\rm \hi}$.
  In the bottom panel we show the ionized-to-neutral ratio $N_{\rm
  \hw}$/$N_{\rm \hi}$ vs $N_{\rm \hi}$.
  The dashed line shows the median value of $N_{\rm \hw}^{\rm Hot}$/$N_{\rm
  \hi}$ from our nine intervening DLAs. 
  Data points from the three DLAs at $<$5\,000~\kms\ from the
  QSO are highlighted in large open circles.
} 
\end{figure}

\section{Contribution of DLA plasma to the cosmic baryon and
  metal density}
The contribution of \hi\ in DLAs to the cosmic density can be
calculated using
\begin{equation}
\Omega^{\rm \hi}({\rm DLA})=\frac{H_0\mu m_{\rm H}}
{c\rho_{\rm crit}}\int_{10^{20.3}}^{10^{21.5}}
N_{\rm \hi}f(N_{\rm \hi}){\rm d}N_{\rm \hi},
\end{equation}
where $f(N_{\rm \hi})$ is the \hi\ column density distribution function.
$\Omega^{\rm \hi}$ has been calculated as $\approx1.0\times10^{-3}$,
fairly flat with redshift \citep*{Pr05, Pe03, Pe05}. 
In Eqn. (5) the lower integration limit corresponds
to the definition of a DLA, and the upper integration limit
corresponds to the break in the power law reported by \citet{Pr05};
above log\,$N_{\rm \hi}$=21.5 $f(N_{\rm \hi})$ becomes much steeper and the
contribution of DLAs at these column densities to $\Omega^{\rm \hi}$
becomes small. 
Since $f$(\hi) has a power law dependence with slope $\alpha\approx-2$ 
\citep{Pr05}, it can be shown by integrating Eqn. (5) that equal bins
of log\,$N_{\rm \hi}$ contain equal contributions to $\Omega$,
i.e. all DLAs contribute toward the mass density. 
The contribution of the hot phase of DLA to the closure
density can be expressed in a similar way:
\begin{equation}
\Omega^{\rm \hw}({\rm DLA})=\frac{H_0\mu m_{\rm H}}
{c\rho_{\rm crit}}\int_{10^{20.3}}^{10^{21.5}}
\frac{N_{\rm \hw}}{N_{\rm \hi}}N_{\rm
  \hi}f(N_{\rm \hi}){\rm d}N_{\rm \hi}. 
\end{equation}

The ratio $N_{\rm \hw}/N_{\rm \hi}$ may depend on $N_{\rm \hi}$, but
since we only have nine intervening DLAs, we do not have enough data
to characterize this relationship. Therefore we proceed by using our
median values 
$\langle N_{\rm \hw}^{\rm Hot}\,/N_{\rm \hi}\rangle>0.4$ and 
$\langle N_{\rm \hw}^{\rm Warm}\,/N_{\rm \hi}\rangle>0.2$
measured over the nine intervening DLAs with \os\footnote{Using the
  mean rather 
than the median raises the $N_{\rm \hw}^{\rm Hot}/N_{\rm \hi}$ ratio
to $\approx$1. However we use the median since it is less sensitive to the
effect of false \os\ detections.}.
Now the contribution from the ionized gas simply becomes 

\begin{equation}
\Omega^{\rm \hw}({\rm DLA})=\langle N_{\rm \hw}/N_{\rm \hi} \rangle
\Omega^{\rm \hi}({\rm DLA}),
\end{equation}

evaluating to $\Omega^{\rm Hot\,\hw}({\rm DLA})>4\times10^{-4}$ and
  $\Omega^{\rm Warm\,\hw}({\rm DLA})>2\times10^{-4}$. 
The contribution of the \emph{metals} in DLAs 
$\Omega_{\rm Z}^{\rm\hi}({\rm DLA})=\langle{\rm Z/H}\rangle\Omega^{\rm
  \hi}({\rm DLA})=4\times10^{-7}$ \citep{Fe05}
then needs to be augmented by 
$\Omega_{\rm Z}^{\rm Hot\,\hw}({\rm DLA})>2\times10^{-7}$
and $\Omega_{\rm Z}^{\rm Warm\,\hw}({\rm DLA})>1\times10^{-7}$. 

Note that the plasma contributions to the baryon and metal budgets would rise
if the value of $f_{\rm \os}$ in the DLA plasma were $<$0.2.
The total density of metals estimated by integrating
the star formation history of the Universe up to $z=2$ is
$\Omega_{\rm Z}^{\rm SFH}\approx3\times10^{-5}$ \citep[using the star
  formation history from][ and assuming a metal yield $y$=1/42]{Bo04}.
Stars in galaxies appear to contain $\approx20$\% of the total
\citep{Bo06}, the contribution from the ISM in galaxies (\hi\ in DLAs)
is $\sim$1\%, and the IGM contains a further $\approx5-25$\% 
\citep{Pe04, Be05}.
The remaining metals ($\approx50$\% of the total) are yet to be
found, hence the formulation of the ``missing metal problem''
\citep*{Pe99, Fe05}. 
If $f_{\rm \os}$ in the DLA plasma were as low as $3\times10^{-3}$, 
which is the case for plasma in CIE at $10^6$~K \citep{SD93, GS06}, then
the \os-bearing plasma around DLAs would constitute 
$\Omega_{\rm Z}^{\rm Hot\,\hw}=1.5\times10^{-5}$ 
and hence solve the missing metals problem. 
In this event the majority of oxygen atoms would be ionized up to
\ion{O}{vii} and \ion{O}{viii}. The lines of these ions are in the
X-ray, but unfortunately searches for high-redshift DLA absorption in
\ion{O}{vii} and \ion{O}{viii} to confirm the presence of $10^6$~K plasma 
are beyond the capabilities of current X-ray satellites.
A scenario with \os\ existing as a trace ionization state in
million-degree DLA coronae is consistent with the broader \os\
lines in our sample, since $b_{\rm \os}=30$~\kms\ in a thermally
broadened line at $10^6$\,K. 
However, this scenario is not required by the data; it is merely one possible
explanation. Indeed, the notion that all the missing metals are in hot
DLA halos is implausible, since metals will also be found in both the
neutral and ionized phases of other classes of QAL system. 
This includes the plasma in sub-DLAs and Lyman Limit Systems
\citep{KT97, KT99, DP01, Pr06}, which
may also probe galactic halos \citep[see Figure 1 in][]{Ma03}, and
could easily contribute to the missing metals.
It will be necessary to search for and
characterize the \os\ phase in QSO absorbers over all ranges of
$N_{\rm \hi}$ to fully measure the quantity of
baryons and metals hidden in hot galactic halos at high redshift.

\section{Discussion}
A full discussion of models of the origin of high-ionization plasma in
DLAs \citep[e.g.][]{M196, Mo96, Ka96, Ra97, WP00b, Ma03} is beyond the
scope of this 
paper. However, we briefly mention two leading theories in light of
the findings in our new data. 

\subsection{Accretion}
Numerical simulations of cosmological structure formation \citep{CO99,
Da01} predict that gas falling onto diffuse large-scale structures
will be accelerated to supersonic speeds and become shock-heated to
temperatures of 10$^5$ to 10$^7$~K, creating a phase known as the warm-hot IGM
(WHIM).\footnote{Note that the term ``warm-hot'' here is at odds with
  our use of the terms warm and hot throughout this paper.}
The \citet{Da01} WHIM evolution models predict $<$10\% of
the cosmic baryons are in the WHIM at $z=3$ (at over-densities
$10<\rho/\bar{\rho}<30$), but 30--40\% are at $z=0$.
Our data show that collisionally ionized plasma at $T>10^5$\,K
exists around DLAs at $z\ga2$, and that the baryonic content of the plasma
is at least of similar order to that in the \hi\ phase. 
This plasma could be identified with the
high-density portions of the WHIM,
since DLAs presumably trace deep potential wells within dark matter halos,
and hence represent likely sites for accretion to occur.
However, the presence of metals in DLA halos
implies that the gas has been processed by star formation; 
accretion would bring in pristine, low-metallicity material.

\subsection{Supernovae and galactic winds}
Star formation in DLAs will lead to supernovae and regions of hot,
shock-heated plasma, which will become outflows if enough energy is
injected. Galactic outflows are a topic of considerable interest, not
least because of the metal pollution and energy injection they bring
about in the IGM \citep[e.g.][]{Ve05, Ag01, Ag05, Ar04}. 

Our DLA observations show \os\ and \cf\ components displaced from the
neutral gas by up to 400~\kms; for comparison, \citet{Ag01} find that
300~\kms\ winds at $z>3$ can account for the metallicity of the Lyman
alpha forest. 
Our data also show evidence at the 98\% level for both a [Z/H]$_{\rm Neut}$ vs
$N_{\rm \os}$ correlation and a [Z/H]$_{\rm Neut}$ vs $\Delta v_{\rm
  \cf}$ correlation. 
These correlations are consistent with a feedback scenario where
supernovae in DLA disks produce both metals and kinematically
disturbed volumes of interstellar plasma.

\section{Summary}
We have performed a successful search for \os\ and other high ions in a
VLT/UVES sample of 35 DLAs with data covering \os\ at S/N$>$10, and
studied the properties of the highly ionized phase and its relationship
to the neutral gas. We now list the principal observational results.

\begin{enumerate}

\item We detect \nos\ cases of DLA \os\ absorption, of which three 
  are at $<$5\,000~\kms\ from the QSO.
  No \os\ non-detections are found, but considering the difficulty in
  making non-detections, a conservative estimate of the incidence of \os\
  absorption in DLAs is $>$34\%.
  All the DLAs show \cf\ and \sif\ when data is available, and
  \nf\ is seen in 3/9 DLAs with \nf\ coverage.
  The \os\ absorbers are characterized by log\,$N_{\rm \os}$ between 13.66
  and $>$15.25 with a median value of 14.77. 
  These average column densities are 
  similar to the amounts measured in the Galactic halo, LMC, and
  starburst galaxy NGC~1705, even though the DLA metallicities
  (measured in the neutral gas) are typically one fortieth of the
  solar values.

\item The \sif\ and \cf\ profiles are very similar, suggesting these
  two ions are generally co-spatial. 
  Many of the individual \cf\ and \sif\ components are narrow enough
  ($b<10$~\kms) to imply photoionization as the origin mechanism. 
 
\item  In nine of ten cases the 
  $N_{\rm \cf}/N_{\rm \os}$ ratio is 
  non-linear with velocity through the profile,
  implying that \os\ and \cf\ are not fully co-spatial.
  No narrow, photoionized \os\ components are seen in the data.
  Furthermore, CLOUDY photoionization models show no single-phase
  solution to the \sif, \cf, and \os\ column densities.
  Thus, the observations suggest the plasma in DLAs contains two phases: a
  photoionized phase seen in \sif\ and \cf, and a hot phase seen in
  \os\ and \nf\ that may also contain additional amounts of \cf. 

\item \cf\ and \os\ components are seen centered up to 400~\kms\ away from the
  neutral gas, indicating inflows or outflows. The total width of the \cf\
  absorbers is typically twice as large as the total width of the neutral
  absorbers.

\item Our data show tentative evidence (98\% significance) for
  correlations between [Z/H]$_{\rm Neut}$ and log\,$N_{\rm \os}$, 
  and between [Z/H]$_{\rm Neut}$ and $\Delta v_{\rm \cf}$. 
  This can be explained if the DLA plasma is produced as a
  direct result of supernova-generated outflows. 
  The high ion properties are uncorrelated to $N_{\rm \hi}$ in the neutral gas.
 
\item Using the observed $N_{\rm \cf}$ and $N_{\rm \os}$ measurements in the
  DLAs from this paper, the metallicities and $N_{\rm \hi}$
  measurements from \citet{Le06}, conservative assumptions for the
  \os\ and \cf\ ionization fractions
  ($f_{\rm \os}<0.2$ and $f_{\rm \cf}<0.3$), 
  and assuming that the plasma has the same metallicity as the
  neutral phase,
  we calculate the total hydrogen column density in the \os-bearing
  and \cf-bearing phases. We find
  log\,$N_{\rm \hw}^{\rm Hot}$ ranges from $>$19.5 to $>$21.1, and 
  log\,$N_{\rm \hw}^{\rm Warm}$ ranges from $>$19.4 to $>$20.9. 
  These column densities of ionized plasma are typically of the same
  order as log\,$N_{\rm \hi}$ in the neutral phase. 
  Using the median 
  value to $N_{\rm \hw}^{\rm Hot}/N_{\rm \hi}$, we find that 
  the total baryonic content of the \os\ phase is
  $\Omega^{\rm Hot\,\hw}({\rm DLA})>4\times10^{-4}$ 
  and the total metal content is
  $\Omega_{\rm Z}^{\rm Hot\,\hw}({\rm DLA})>2\times10^{-7}$, 
  i.e. $>$1\% of the metal budget at $z\approx2$.
  If the temperature in the \os-bearing plasma is $\approx10^6$\,K and so
  $f_{\rm \os}\ll 0.2$, then the \os\ phase of DLAs can make a significant
  contribution to the missing metals problem.
\end{enumerate}
 
{\bf Acknowledgements}\\
AJF is supported by a Marie Curie Intra-European Fellowship awarded by
the European Union Sixth Framework Programme.
We are grateful to Philipp Richter and Blair Savage for comments,
to Bart Wakker for help in implementing CLOUDY,
and to Jacqueline Bergeron and Patrick Boiss\'e for assistance with VPFIT.
PP and RS gratefully acknowledge support from the Indo-French
Centre for the Promotion of Advanced Research (Centre Franco-Indien
pour la Promotion de la Recherche Avanc\'ee) under contract No. 3004-3.
We thank the referee for a thorough report that improved the
quality of the manuscript.

\clearpage
\small
\begin{table*}
\caption{High Ion Column Densities in DLAs}
\begin{minipage}[t]{16cm}
\centering
\renewcommand{\footnoterule}{}
\begin{tabular}{lcccc cccc}
\hline\hline
QSO\footnote{Alternate (SIMBAD-compatible) names are given in parentheses.} & $z_{\rm qso}$\footnote{Redshift of quasar determined from position of \lya\ emission.} & $z_{\rm abs}$ & Ion & $v_-$\footnote{$v_-$ and $v_+$ are the lower and upper velocity limits of absorption in \kms.} & $v_+\,^{c}$ & log\,$N_{\rm{a}}$(strong)\footnote{Column density measured using AOD method. Strong and weak refer to the two lines of the doublet.} & log\,$N_{\rm{a}}$(weak)$^{d}$ & log\,$N$(fit)\footnote{Column density measured using VPFIT. The number in parentheses indicates the number of Voigt components in the model.}\\
\hline
        \object{Q0027-186} & 2.56 &  2.402 &       \os &  $-$100 &      60                             \footnote{AOD measurement made over a limited velocity range, due to partial blending. $N$(fit), determined using unblended data from both lines, may be higher.}       & ...\footnote{Line completely blended.} & 
14.81$\pm$0.03 & 
15.09$\pm$0.03( 5) \\
                           &       &         &  \sif &  $-$100 &     250     & 
13.77$\pm$0.03 & 
13.66$\pm$0.04 & 
13.74$\pm$0.03( 5) \\
                           &       &         &   \cf &  $-$100 &     250     & 
$>$14.67 & 
$>$14.68 & 
14.69$\pm$0.03( 9) \\
        \object{Q0112+306} & 2.99 &  2.702 &       \os &  $-$100 &     195$^f$                                      & ...$^g$ & 
$>$15.15 & 
15.21$\pm$0.05( 3) \\
                           &       &         &  \sif &  $-$230 &     225     & 
14.22$\pm$0.03 & 
14.24$\pm$0.03 & 
14.23$\pm$0.09(12) \\
                           &       &         &   \cf &  $-$230 &     225     & 
$>$14.87 & 
14.83$\pm$0.03 & 
14.84$\pm$0.03( 8) \\
        \object{Q0450-131} &  2.25 &   2.067 &   \os &  $-$100 &     170     & 
$>$15.12 & 
$>$15.12 & 
15.12$\pm$0.13( 4) \\
(\object{QSO B0450-1310B}) &       &         &   \nf &  $-$ 50 &     110     & 
13.56$\pm$0.03 & 
13.59$\pm$0.03 & 
13.57$\pm$0.03( 3) \\
        \object{Q0528-250} &  2.77 &   2.811 &   \os &  $-$200 &      80$^f$ & 
$>$15.14 & 
...$^g$ &
15.18$\pm$0.03( 4) \\
                           &       &         &  \sif &  $-$260 &     300     & 
$>$14.45 & 
$>$14.58 & 
14.57$\pm$0.03(13) \\
                           &       &         &   \cf &  $-$260 &     140     & 
$>$14.97 & 
$>$15.09 & 
15.29$\pm$0.09( 6) \\
                           &       &         &   \nf &  $-$150 &     150     & 
14.02$\pm$0.03 & 
14.00$\pm$0.03 & 
14.04$\pm$0.03( 4) \\
        \object{Q0841+129} &  2.50 &   2.476 &   \os &  $-$ 90 &  $-$  5$^f$ & 
13.91$\pm$0.04 & 
13.94$\pm$0.05 & 
13.95$\pm$0.09( 2) \\
                           &       &         &   \nf &  $-$ 85 &       0     & 
$<$12.59 & 
...$^g$ &
... \\
        \object{Q0913+072} &  2.78 &   2.618 &   \os &  $-$ 22 &      80$^f$ & 
14.43$\pm$0.03 & 
14.43$\pm$0.03 & 
14.49$\pm$0.04( 2) \\
                           &       &         &  \sif &  $-$ 35 &     220     & 
13.64$\pm$0.03 & 
13.69$\pm$0.03 & 
13.67$\pm$0.03( 5) \\
                           &       &         &   \cf &  $-$ 35 &     220     & 
14.13$\pm$0.03 & 
14.14$\pm$0.03 & 
14.14$\pm$0.03( 5) \\
        \object{Q1337+113} &  2.92 &   2.796 &   \os &  $-$100 &      20     & 
13.66$\pm$0.06 & 
...$^g$ &
13.66$\pm$0.03( 1) \\
                           &       &         &   \cf &  $-$100 &      25     & 
13.37$\pm$0.05 & 
13.38$\pm$0.07 & 
13.40$\pm$0.03( 1) \\
                           &       &         &   \nf &  $-$100 &      25     & 
$<$12.29 & 
...$^g$ &
... \\
        \object{Q1409+095} & 2.85 &  2.456 &       \os &  $-$ 35 &      70                                          & ...$^g$ & 
14.27$\pm$0.03 & 
14.28$\pm$0.03( 2) \\
                           &       &         &  \sif &  $-$ 35 &      50     & 
12.97$\pm$0.03 & 
12.93$\pm$0.03 & 
12.97$\pm$0.03( 2) \\
                           &       &         &   \cf &  $-$ 55 &      70     & 
13.74$\pm$0.03 & 
13.74$\pm$0.03 & 
13.74$\pm$0.03( 3) \\
                           &       &         &   \nf &  $-$ 20 &      85     & 
$<$12.75 & 
...$^g$ &
... \\
        \object{Q2059-360} & 3.09 &  3.083 &       \os &  $-$ 60 &     125                                          & ...$^g$ & 
14.07$\pm$0.04 & 
14.08$\pm$0.03( 1) \\
                           &       &         &  \sif &  $-$ 60 &     125     & 
12.87$\pm$0.07 & 
...$^g$ &
12.93$\pm$0.03( 2) \\
                           &       &         &   \cf &  $-$ 60 &     125     & 
13.73$\pm$0.03 & 
13.76$\pm$0.03 & 
13.73$\pm$0.03( 3) \\
                           &      &        &       \nf &  $-$ 60 &     125                                          & ...$^g$ & 
$<$13.02 & 
... \\
        \object{Q2138-444} &  3.17 &   2.852 &   \os &  $-$300 &      20$^f$ & 
14.77$\pm$0.03 & 
...$^g$ &
14.78$\pm$0.03( 3) \\
                           &       &         &   \cf &  $-$290 &     200     & 
14.45$\pm$0.03 & 
14.43$\pm$0.03 & 
14.45$\pm$0.03(13) \\
                           &      &        &       \nf &  $-$290 &     200                                          & ...$^g$ & 
$<$13.06 & 
... \\
        \object{Q2206-199} &  2.56 &   2.076 &   \os &  $-$ 27 &      38     & 
13.76$\pm$0.03 & 
13.81$\pm$0.04 & 
13.84$\pm$0.03( 1) \\
 (\object{LBQS2206-1958A}) &      &        &      \sif &  $-$ 35 &      50                                          & ...$^g$ & 
12.82$\pm$0.03 & 
12.85$\pm$0.03( 2) \\
                           &       &         &   \cf &  $-$ 35 &      50     & 
13.71$\pm$0.03 & 
13.72$\pm$0.03 & 
13.71$\pm$0.03( 2) \\
                           &      &        &       \nf &  $-$ 35 &      50                                          & ...$^g$ & 
$<$12.75 & 
... \\
        \object{Q2243-605} &  3.01 &   2.331 &   \os &  $-$220 &  $-$ 80$^f$ & 
$>$14.58 & 
...$^g$ &
14.98$\pm$0.03( 3) \\
    (\object{HE2243-6031}) &       &         &  \sif &  $-$400 &  $-$ 13     & 
$>$13.88 & 
...$^g$ &
14.37$\pm$0.42( 8) \\
                           &       &         &   \cf &  $-$480 &      15     & 
$>$14.68 & 
$>$14.73 & 
14.74$\pm$0.05(14) \\
                           &       &         &   \nf &  $-$380 &  $-$335     & 
13.47$\pm$0.03 & 
13.50$\pm$0.03 & 
13.56$\pm$0.03( 2) \\
\hline
\end{tabular}
\end{minipage}
\end{table*}
\normalsize

\begin{table*}
\begin{minipage}[t]{16cm}
\caption{DLA kinematics: neutral vs \cf\ vs \os}
\centering
\renewcommand{\footnoterule}{}
\begin{tabular}{lcccc ccc}
\hline \hline
QSO & $z_{\rm abs}$\footnote{Redshift of the system determined using the velocity centroid of the strongest component in the neutral line.}  & Neutral  & $\Delta v_{\rm neut}$\footnote{$\Delta v$ is the velocity range enclosing the central 90\% of the total integrated optical depth in the line (in \kms).}& $\bar{v}_{\rm \cf}$\footnote{$\bar{v}$ is the optical-depth weighted mean velocity of the line profile, $\bar{v}=\int^{v_+}_{v_-}v\tau_a(v){\rm d}v/\int^{v_+}_{v_-}\tau_a(v){\rm d}v$ (in~\kms).} & $\Delta v_{\rm \cf}^b$ & $\bar{v}_{\rm \os}^c$ & $\Delta v_{\rm \os}^b$\\
\hline
     \object{Q0027-186} &   2.402 &  \ion{Si}{ii}~1526 & 180$\pm$2 &                                                      104$\pm$4 & $\la$227\footnote{Lines saturated, so $\Delta v$ could be lower.} &                                         ...\footnote{Lines partly blended and saturated, so no measurement is possible.} & ...$^e$ \\
     \object{Q0112+306} &   2.702 &  \ion{Fe}{ii}~1608 & 208$\pm$2 &                                                                                                         $-$22$\pm$2 & $\la$295$^d$ &         ...\footnote{Lines partly blended, so $\bar{v}$ is unknown and only a lower limit to $\Delta v$ is measured.} & $>$240$^f$ \\
     \object{Q0450-131} &   2.067 &  \ion{Fe}{ii}~1608 & 134$\pm$2 &                                                                                    ...\footnote{No available \cf\ data.} & ...$^g$ &                                                                                                            20$\pm$4 & $\la$168$^d$ \\
     \object{Q0528-250} &   2.811 &   \ion{S}{ii}~1253 & 282$\pm$2 &                                                                                                            21$\pm$4 & $\la$363$^d$ &                                                                                                                  ...$^e$ & ...$^e$ \\
     \object{Q0841+129} &   2.476 &   \ion{S}{ii}~1259 &  26$\pm$2 &                                                                                                                  ...$^g$ & ...$^g$ &                                                                                                               ...$^f$ & $>$ 67$^f$ \\
     \object{Q0913+072} &   2.618 &  \ion{Si}{ii}~1526 &  24$\pm$2 &                                                                                                               78$\pm$2 & 175$\pm$2 &                                                                                                               ...$^f$ & $>$ 88$^f$ \\
     \object{Q1337+113} &   2.796 &  \ion{Fe}{ii}~1608 &  36$\pm$2 &                                                                                                            $-$28$\pm$4 &  80$\pm$5 &                                                                                                            $-$27$\pm$7 &  69$\pm$9 \\
     \object{Q1409+095} &   2.456 &  \ion{Si}{ii}~1526 &  65$\pm$2 &                                                                                                               17$\pm$2 &  74$\pm$2 &                                                                                                               22$\pm$2 &  94$\pm$2 \\
     \object{Q2059-360} &   3.083 &  \ion{Fe}{ii}~1121 &  44$\pm$2 &                                                                                                               40$\pm$2 & 103$\pm$2 &                                                                                                               24$\pm$4 &  97$\pm$6 \\
     \object{Q2138-444} &   2.852 &  \ion{Si}{ii}~1808 &  44$\pm$2 &                                                                                                            $-$65$\pm$2 & 342$\pm$2 &                                                                                                               ...$^f$ & $>$249$^f$ \\
     \object{Q2206-199} &   2.076 &  \ion{Si}{ii}~1304 &  21$\pm$2 &                                                                                                                5$\pm$2 &  47$\pm$2 &                                                                                                               10$\pm$2 &  42$\pm$2 \\
     \object{Q2243-605} &   2.331 &  \ion{Si}{ii}~1808 & 165$\pm$2 &                                                                                                        $-$270$\pm$6 & $\la$344$^d$ &                                                                                                                  ...$^e$ & ...$^e$ \\
\hline
\multicolumn{3}{l}{{\bf Mean and std dev\footnote{Upper limits treated as data points in these calculations.}}}  & 
102$\pm$88 & ... & 205$\pm$125 & ... & ...\\
\hline
\end{tabular}
\end{minipage}
\end{table*}

\begin{table*}
\begin{minipage}[t]{16cm}
\caption{Total hydrogen column densities in \os\ phase, \cf\ phase, and neutral phase} 
\centering
\renewcommand{\footnoterule}{}
\begin{tabular}{lcccc ccccc}
\hline \hline
QSO & $z_{\rm abs}$ & [Z/H]$_{\rm Neut}$\footnote{[Z/H]$_{\rm Neut}$ and log\,$N_{\rm \hi}$ from \citet{Le06}. Element Z is Zn if \ion{Zn}{ii} is detected, otherwise S or Si.} & log\,$N_{\rm \hi}^a$ & log\,$N_{\rm \os}$ & log\,$N_{\rm \hw}^{\rm Hot}$\footnote{$N_{\rm \hw}^{\rm Hot}=N_{\rm \os}/[f_{\rm \os}{\rm (O/H)_{Hot}}]$; we present a lower limit since $f_{\rm \os}<0.2$. We assume [O/H]$_{\rm Hot}$=[Z/H]$_{\rm Neut}$.} & $N_{\rm \hw}^{\rm Hot}/N_{\rm \hi}$ & log\,$N_{\rm \cf}$ & log\,$N_{\rm \hw}^{\rm Warm}$\footnote{$N_{\rm \hw}^{\rm Warm}=N_{\rm \cf}/[f_{\rm \cf}{\rm (C/H)_{Warm}}]$; we present a lower limit since $f_{\rm \cf}<0.3$. We assume [C/H]$_{\rm Warm}$=[Z/H]$_{\rm Neut}$.} & $N_{\rm \hw}^{\rm Warm}/N_{\rm \hi}$\\
\hline
    \object{Q0027-186} &   2.402 & 
   $-$1.63$\pm$0.10 & 21.75$\pm$0.10 & 
15.09$\pm$0.03 & 
$>$20.8 & $>$0.10  &
$>$14.68 & 
$>$20.4 & $>$0.05 \\
    \object{Q0112+306} &   2.702 & 
   $-$0.49$\pm$0.11 & 20.30$\pm$0.10 & 
$>$15.15 & 
$>$19.7 & $>$0.24  &
$>$14.83 & 
$>$19.5 & $>$0.14 \\
    \object{Q0450-131} &   2.067 & 
   $-$1.62$\pm$0.08 & 20.50$\pm$0.07 & 
$>$15.12 & 
$>$20.8 & $>$1.92  &
...&...&...\\
    \object{Q0528-250} &   2.811 & 
   $-$0.91$\pm$0.07 & 21.35$\pm$0.07 & 
$>$15.14 & 
$>$20.1 & $>$0.06  &
$>$15.09 & 
$>$20.1 & $>$0.06 \\
    \object{Q0841+129} &   2.476 & 
   $-$1.60$\pm$0.10 & 20.80$\pm$0.10 & 
$>$13.91 & 
$>$19.5 & $>$0.06  &
...&...&...\\
    \object{Q0913+072} &   2.618 & 
   $-$2.59$\pm$0.10 & 20.35$\pm$0.10 & 
$>$14.43 & 
$>$21.1 & $>$5.07  &
14.13$\pm$0.03 & 
$>$20.9 & $>$3.18 \\
    \object{Q1337+113} &   2.796 & 
   $-$1.86$\pm$0.09 & 21.00$\pm$0.08 & 
13.66$\pm$0.06 & 
$>$19.6 & $>$0.04  &
13.37$\pm$0.05 & 
$>$19.4 & $>$0.02 \\
    \object{Q1409+095} &   2.456 & 
   $-$2.06$\pm$0.08 & 20.53$\pm$0.08 & 
14.27$\pm$0.03 & 
$>$20.4 & $>$0.69  &
13.74$\pm$0.03 & 
$>$19.9 & $>$0.25 \\
    \object{Q2059-360} &   3.083 & 
   $-$1.77$\pm$0.09 & 20.98$\pm$0.08 & 
14.07$\pm$0.04 & 
$>$19.9 & $>$0.08  &
13.73$\pm$0.03 & 
$>$19.6 & $>$0.04 \\
    \object{Q2138-444} &   2.852 & 
   $-$1.74$\pm$0.05 & 20.98$\pm$0.05 & 
$>$14.77 & 
$>$20.5 & $>$0.37  &
14.45$\pm$0.03 & 
$>$20.3 & $>$0.22 \\
    \object{Q2206-199} &   2.076 & 
   $-$2.32$\pm$0.05 & 20.44$\pm$0.05 & 
13.76$\pm$0.03 & 
$>$20.1 & $>$0.48  &
13.71$\pm$0.03 & 
$>$20.2 & $>$0.53 \\
    \object{Q2243-605} &   2.331 & 
   $-$0.85$\pm$0.05 & 20.65$\pm$0.05 & 
$>$14.98 & 
$>$19.9 & $>$0.17  &
$>$14.73 & 
$>$19.7 & $>$0.12 \\
\hline
\end{tabular}
\end{minipage}
\end{table*}

\clearpage 

\appendix
\section{Notes on individual systems}

\indent{\bf Q0027-186 $z_{\rm qso}=2.55$ $z_{\rm abs}=2.402$}\\ 
\os~$\lambda$1031 is blended below $-$50~\kms\ and above 200~\kms, but
shows strong absorption components near 0, 50, 125, and 175~\kms, as
does the \cf\ profile.
\os~$\lambda$1037 appears to be unblended in the ranges 
$-$100 to 60~\kms\ and 180 to 220~\kms, where the
profile shows similar components to those seen in \cf.
In our analysis we adopt log\,$N_{\rm \os}$=15.09$\pm$0.03, 
based on the results from a Voigt fit to the unblended
parts of each \os\ profile.
The AOD integration of \sif~$\lambda$1393 yields a slightly higher
column density (by 0.11~dex) than $\lambda$1402, so
$\lambda$1393 may be partly blended.

{\bf Q0112-306 $z_{\rm qso}=2.99$ $z_{\rm abs}=2.702$}\\ 
Both \os\ $\lambda$1031 and $\lambda$1037 are blended at
$v<-100$~\kms. At $v>-100$~\kms\ both \os\ lines show the same
profile, which is clearly smoother than the \cf.
We only present a lower limit to $N_{\rm \os}$ in this system using a
Voigt fit to the data at $v>-100$~\kms. 
At least eight components are seen in the \cf\ and \sif\ data. 

{\bf Q0450-131 $z_{\rm qso}=2.25$ $z_{\rm abs}=2.067$}\\ 
Both \nf~$\lambda$1238 and $\lambda$1242 show three components.
Strong corresponding \os\ is seen in
$\lambda$1031 and $\lambda$1037, which have consistent optical depth
profiles. \cf\ and \sif\ data do not exist.
The QSO is 17\,900 \kms\ away from the DLA.
Full details of the neutral gas in this system are published in \citet{DZ06}.

{\bf Q0528-250 $z_{\rm qso}=2.77$ $z_{\rm abs}=2.811$}\\ 
This DLA is 3\,600 \kms\ more redshifted than the QSO and may be a QSO
intrinsic absorber. Though the \os\ continuum placement is
difficult due to a high density of lines,
absorption features appear in $\lambda$1031 at 
$-$150, $-$50, and 25~\kms, the velocities of
strong \sif\ and \cf\ absorption components.
Our value for $N_{\rm \os}$ derives from a four component fit to the data.
The data show two very good detections of \nf\ with 
identical column densities measured in each line.

{\bf Q0841+129 $z_{\rm qso}=2.50$ $z_{\rm abs}=2.476$}\\ 
This system is separated from the host quasar by 2050~\kms.
Though there are several blends near \os\ $\lambda$1031 and
$\lambda$1037, two components appears in both lines with the
same optical depth profile in the range $-$100 to 0~\kms.
with the stronger component at $-$65~\kms.
No \cf\ or \sif\ data are currently available, and \nf\ is not
detected. 
The neutral gas in this DLA is studied in \citet{DZ06}.

{\bf Q0913+072 $z_{\rm qso}=2.78$ $z_{\rm abs}=2.618$}\\ 
There are several blends near \os\ $\lambda$1031 and
$\lambda$1037, but we verify \os\ absorption in the velocity range
$-$22 to 80 \kms\ since the $\lambda$1037 profile shows the same
column density profile as $\lambda$1031 in this interval.
The \cf\ and \sif\ data show at least
four components over an interval of $>$200~\kms.

{\bf Q1337+113 $z_{\rm qso}=2.92$ $z_{\rm abs}=2.796$}\\ 
\os\ $\lambda$1037 is blended, but the $\lambda$1031 profile shows a
single component with a similar
profile to \cf, giving a $N_{\rm \cf}/N_{\rm \os}$
ratio that is, within the errors, flat with velocity over a 60~\kms\ range.
This is the only DLA in our sample where this ratio is flat.
No \nf\ is seen.

{\bf Q1409+095 $z_{\rm qso}=2.85$ $z_{\rm abs}=2.456$}\\ 
\os\ $\lambda$1031 is blended, but the $\lambda$1037 profile shows the
same two components 
(central velocities of $-$19 and 35 \kms) 
as are seen in \cf\ and \sif.
In \cf\ and \os, the 35~\kms\ component is stronger than the
$-$19~\kms\ component.
In \sif\ and \siw, the $-$19~\kms\ and 35~\kms\
components are of equal strength, showing that the \sif\ 
follows the neutral gas in this case. 
\nf\ is not detected.

{\bf Q2059-360 $z_{\rm qso}=3.09$ $z_{\rm abs}=3.083$}\\ 
This DLA is only 500~\kms\ away from the QSO. 
\os\ $\lambda$1031 is blended; \os\ $\lambda$1037 shows a clean
detection of a single component, with a similar width to the \cf\ absorption.
Three components appear in the higher S/N \cf\ data, so the \cf/\os\
ratio is non-linear with velocity.
This represents our highest redshift DLA \os\ detection. 

{\bf Q2138-444 $z_{\rm qso}=3.17$ $z_{\rm abs}=2.852$}\\ 
\os\ $\lambda$1031 shows absorption between $-$300 and 0~\kms,
but is blended above 0~\kms. The \os\ $\lambda$1037 profile 
is identical to $\lambda$1031 in the range $-$160 to 20~\kms, 
confirming the detection of \os, though the continuum placement is
difficult.
The \cf\ profiles broadly follows the \os\ $\lambda$1031 profile
in the range $-$300 to 0~\kms, although more subcomponents appear in
the \cf\ data.

{\bf Q2206-199 $z_{\rm qso}=2.56$ $z_{\rm abs}=2.076$}\\ 
This sight line shows a single \os\ absorption component in both
$\lambda$1031 and $\lambda$1037. 
Two components are revealed in the higher S/N \sif\ and \cf\ data,
separated by 20~\kms.
The \os\ appears to be aligned with the weaker \cf\ component
centered at 13~\kms. 
See \citet{PW97a} for a detailed analysis of this DLA; our total \cf\ and
\sif\ column densities are in agreement with the earlier published values.

{\bf Q2243-605 $z_{\rm qso}=3.01$ $z_{\rm abs}=2.331$}\\ 
\os\ $\lambda$1031 and $\lambda$1037 are each blended at different
velocities but \os\ components centered near $-$350, $-$200,
and $-$100~\kms\ correspond to features seen in \cf.
A Lyman-$\gamma$ line at $z=2.531$ is identified at $-$270~\kms\ in
the \os\ $\lambda$1031 frame.
We only present a lower limit to $N_{\rm \os}$ in this system based on
a 3-component Voigt profile fit 
to the unblended data.
\nf\ is detected in both $\lambda$1238 and $\lambda$1242, at velocities
where the \cf\ and \os\ optical depths are highest.
At least 14 components are seen in the \cf\ data over a velocity range
of $\approx$500~\kms. 
\cf\ and \sif\ follow each other in the series of
absorption components in the range $-$225 to 0~\kms, but the \cf\ profile
shows many additional components in the range $-$500 to $-$225~\kms\
that have no counterparts in \sif. 
See \citet{Lo02} for a detailed analysis of this DLA.

\end{document}